\documentclass[a4paper,10pt,aps,prb,nofootinbib]{revtex4-1}

\usepackage[sumlimits,intlimits]{amsmath}
\usepackage{amsfonts}
\usepackage{amssymb}
\usepackage{graphicx}
\usepackage{hyperref}
\usepackage{subfig}
\usepackage{ifthen}
\usepackage{fancybox}

\newboolean{ArXiv}
\setboolean{ArXiv}{true}
\ifthenelse{\boolean{ArXiv}}{\typeout{ArXiv=yes}}{\typeout{ArXiv=no}}  
\newboolean{paper}
\setboolean{paper}{false}
\ifthenelse{\boolean{paper}}{\typeout{paper=yes}}{\typeout{paper=no}}  

\newenvironment{MyShadowBox}{\begin{Sbox}\begin{minipage}}{\end{minipage}\end{Sbox}\shadowbox{\TheSbox}}

\newcommand {\nn}[1]   
	    {\ifthenelse{\boolean{ArXiv}}
            {{	
		\begin{center}
		  \begin{MyShadowBox}{1.0\textwidth}
\centerline{ArXiv}
		    {\sf #1}
		  \end{MyShadowBox}
		\end{center}
	    }}    
	    { }}

\newcommand {\pp}[1]   
	    {\ifthenelse{\boolean{paper}}
            {{	
		\begin{center}
		  \begin{MyShadowBox}{1.0\textwidth}
\centerline{paper}
		    {\sf #1}
		  \end{MyShadowBox}
		\end{center}
	    }}    
	    { }}

\renewcommand {\nn}[1]   
	    { \ifthenelse{\boolean{ArXiv}}{{ #1 }}{{ }} }

\renewcommand {\pp}[1]   
	    { \ifthenelse{\boolean{paper}}{{ #1 }}{{ }} }

\newcommand {\spazio}               {\;\;\;\;\;}
\newcommand {\punto}                {\spazio .}
\newcommand {\virgola}              {\spazio ,}
\newcommand {\puntoevirgola}        {\spazio ;}
\newcommand {\afrac}[2]             {\,\cfrac{#1}{#2}\,}
\newcommand {\pton}[1]              {\ensm{\left(  #1 \right)}}
\newcommand {\pqua}[1]              {\ensm{[ #1 ]}}
\newcommand {\pgra}[1]              {\ensm{\left\{ #1 \right\}}}
\newcommand {\ensm}                 {\ensuremath}
\newcommand {\av}[1]                {\ensm{\langle {#1} \rangle}}
\newcommand {\abs}[1]               {\ensm{\left | {#1} \right |}}
\newcommand {\DD}[1]                {\ensm{\mathinner{\Delta\! #1}}}
\newcommand {\ddd}[1]               {\ensm{\mathinner{\delta #1}}}
\newcommand { \lt }                 { < }

\newcommand {\ifoif}                {\ensm{\spazio\Longleftrightarrow\spazio}}
\newcommand {\arrforw}              {\ensm{\spazio\Longrightarrow\spazio}}
\newcommand {\tz}[1]                {\widetilde{#1}}
\newcommand {\est}[1]               {\hat{#1}}
\newcommand {\ve}[1]                {\ensm{\boldsymbol{#1}}}
\newcommand {\thx}                  {\ensm{\theta_x}}
\newcommand {\thy}                  {\ensm{\theta_y}}
\newcommand {\thtx}                 {\ensm{\theta_{\tz{x}}}}
\newcommand {\thty}                 {\ensm{\theta_{\tz{y}}}}
\newcommand {\scaleFact}            {\ensm{\tau}}

\DeclareMathOperator*{\sign}        { \, Sign }
\DeclareMathOperator*{\covariance}  { \, Cov }
\DeclareMathOperator*{\variance}    { \, Var }

\newif\iftest
\ifx\diffl\undefined
	\newcommand{\diffl}[1]                {\mathinner{d#1}}
\else
\fi

\newif\iftest
\ifx\pdiffl\undefined
	\newcommand{\pdiffl}[1]               {\mathinner{\partial#1}}
\else
\fi
\newif\iftest
\ifx\deriv\undefined
	\newcommand{\deriv}[2]      {\cfrac{\diffl{#1}}{\diffl{#2}}}
\else
\fi
\newcommand{\difflsq}[1]              {\mathinner{d^2#1}}
\newif\iftest
\ifx\derivsq\undefined
\newcommand{\derivsq}[2]              {\afrac{\difflsq{#1}}{\diffl{#2}^2}}
\else
\fi
\newif\iftest
\ifx\pderiv\undefined
  \newcommand{\pderiv}[2]               {\cfrac{\pdiffl{#1}}{\pdiffl{#2}}}
\else
\fi

\newcommand {\chiSq}                {\ensm{\chi^2}}

\newcommand {\VarX}                 {\ensm{V_x}}
\newcommand {\VarY}                 {\ensm{V_y}}
\newcommand {\CovXY}                {\ensm{C_{xy}}}
\newcommand {\StDvX}                {\ensm{\sigma_x}}
\newcommand {\StDvY}                {\ensm{\sigma_y}}

\newcommand {\VarXt}                {\ensm{V_{\tz{x}}}}
\newcommand {\VarYt}                {\ensm{V_{\tz{y}}}}
\newcommand {\CovXYt}               {\ensm{C_{\tz{x}\tz{y}}}}
\newcommand {\StDvXt}               {\ensm{\sigma_{\tz{x}}}}
\newcommand {\StDvYt}               {\ensm{\sigma_{\tz{y}}}}

\newcommand {\VarCsi}               {\ensm{V_{\xi }}}
\newcommand {\VarEta}               {\ensm{V_{\eta}}}
\newcommand {\CovXiEta}             {\ensm{C_{\xi\eta}}}

\begin{document}

\nn{
\title{
Least-squares fit to a straight line when each variable contains all equal errors
}

\author{Alessandro Petrolini}
 \affiliation{
   Dipartimento di Fisica dell'Universit\`a di Genova and INFN, 
   Via Dodecaneso 33, I-16146, Genova, Italy.
 }
 \email{alessandro.petrolini@ge.infn.it}   

\date{\today}

\begin{abstract}

The least squares fit to a straight line, when both variables are affected by all
equal uncorrelated errors, leads to very simple results for both the
estimated parameters and their standard errors, of widespread
applicability.  
In this paper several formulas are derived, presenting a full set of results
about the estimated parameters and their standard errors.
All the results have been validated with extensive Monte Carlo simulations.
The emphasis of the paper is on the calculation and properties of the best-fit
parameters and their standard
errors.

\vspace{5 mm}

\begin{center}
This paper is the expanded version of the article entitled \\
\textit{Linear least squares fit when both variables are affected by equal uncorrelated errors}, \\
published in Am. J. Phys. vol. 82, 1178 (2014); \url{http://dx.doi.org/10.1119/1.4893679}.
\end{center}

\end{abstract}

\maketitle
}

\pp{
\title{
Standard errors on the parameters of the least squares fit to a straight line
when both variables have all equal uncorrelated errors
}

\author{Alessandro Petrolini}
 \affiliation{
   Dipartimento di Fisica dell'Universit\`a di Genova and INFN, 
   Via Dodecaneso 33, I-16146, Genova, Italy.
 }
 \email{alessandro.petrolini@ge.infn.it}   

\date{\today}

\begin{abstract}

The least squares fit to a straight line, when both variables are affected by all
equal uncorrelated errors, leads to very simple results for both the
estimated parameters and their standard errors, of widespread
applicability.  
In this paper several formulas are derived, presenting a full set of results
about the estimated parameters and their standard errors.
All the results have been validated with extensive Monte Carlo simulations.
The emphasis of the paper is on the calculation and properties of the best-fit
parameters and their standard
errors.

\end{abstract}

\maketitle
}

\tableofcontents

\section{Introduction}

Least squares fit to a straight line (LSFSL) is the subject of an extensive
literature, not only in the field of physics.
\pp{
Many contributions appeared in this Journal during the last thirty years.
}  
See, for instance, the reviews in~\cite{bi:Macdonald,bi:York} and the references
therein to the older literature.

A good understanding of the LSFSL is important, not only for students but also
for researchers in physical sciences, because least squares fits, nowadays, are
typically done by black-box computer programs.  On the other hand, the
availability of fast computer programs allows one to carry on long and intensive
calculations and Monte Carlo simulations in a short time, often avoiding the use
of approximations.

In physics, problems of LSFSL are often met such that both variables are
affected by significant measurements errors, equal for all measured data points
separately for both variables, and with the errors in two variables
uncorrelated; this is the so-called 
\textit{Standard Weighting Model}~\cite{bi:Macdonald,bi:PHBorcherdstAndCVShetht} (SWM). 
In fact, the problem which triggered the work presented in this paper was the
problem of reconstruction of straight line image tracks in pixelized single-photon detectors
with rectangular pixels, designed for astro-particle physics
experiments~\cite{bi:SEUSO}.
Similar problems are often encountered in high-energy particle physics, when
dealing with the reconstruction of particle tracks, and in the field of imaging by means of
pixelized detectors.
In astrophysics color-color diagrams typically lead to similar problems (see section~\ref{pa:casesReed}).
The method proposed in this paper has been also applied in~\cite{bi:Wehus}.
Moreover the SWM is
often assumed whenever the errors of the data points are unknown and there is no
reason to assume that the errors in one variable are negligible with
respect to the errors in the other one. In this case
the common unknown value of the error can be estimated by the fit.

The LSFSL-SWM problem can be always reduced, by a suitable rescaling of the variables, to an
equivalent problem with the new dimensionless variables having equal
errors~\cite{bi:RBarlow}. The sum of the distances between the measured data points
and the best-fit straight line can thus be minimized leading to a purely
geometrical problem.  
It is possible to solve this problem in an exact way and the results are very simple. 

The purpose if this paper is to present a complete set of explicit 
results for the LSFSL-SWM. The results from the physics literature
known to the author are quickly reviewed, and other new results are derived,
including very simple
analytic formulas for the standard errors of the parameters.
It is shown that parameterizing the straight line with the angle with respect to
one of the axes plus a second dimensional parameter leads to very simple
results, having simple transformation properties under roto-translation of the
Cartesian Coordinate System.
It is shown that variances and covariance of the parameters can always be expressed in
terms of the standard error of the angle plus purely geometrical quantities.
All the results and the accuracy of the expressions 
for the standard errors have been cross-checked with
extensive Monte Carlo simulations.
The results are also compared with the results of the ordinary least squares
(OLS) fit, the case of significant
errors in only one of the variables, which will be called, in short~\cite{bi:Macdonald},
\textit{\mbox{OLS-y:x} fit} (error on the $y$ variable only) and 
\textit{\mbox{OLS-x:y} fit} (error on the $x$ variable only).
A simple criterion for neglecting the error in one of the two variables is derived.
 
Emphasis of this paper is on the analytical, computational and practical
aspects of the problem, not on its rigorous statistical treatment.
\nn{
All the formulas needed for a real implementation in a programming code are derived and discussed.
}

The outline of the paper is as follows.
In section~\ref{pa:Problem} the problem and hypotheses are presented. Results readily available in
the physics literature are summarized in section~\ref{pa:SlopeIntercept}.
All the new results are presented in Section~\ref{pa:AngleAndSignedDistance}.
Section~\ref{pa:cases} presents a few case studies.
Section~\ref{pa:simul} summarizes some of the results obtained by Monte Carlo
simulations carried on both to cross-check the formulas and to evaluate the
accuracy of the standard errors.
\nn{
Finally the appendixes collect all the calculations.
}
\pp{
Many more details, including all mathematical details, and a more detailed
discussion of all the items of this paper
can be found in the companion note~\cite{bi:nota}.
}

\section{Hypotheses and the SWM}
\label{pa:Problem}

For the sake of simplicity let us introduce the symbol for the average of the
generic random quantity $q$, $\av{q}$, the symbol $ \sigma [ q ] $ for its
standard deviation and the symbol $ \rho_{pq} $ for the correlation coefficient
of the two random variables $p$ and $q$.

\pp{

Consider a set of $N$ measured data points, described by the $\tz{x}$ and $\tz{y}$ coordinates
of a suitable Cartesian Coordinate System: $\pgra{ \tz{x}_k ; \tz{y}_k }, ( k = 1, \ldots, N )$.
Assume that the measured data points follow the SWM, that is:
have equal standard errors, separately in both variables, 
$\StDvXt \equiv \sigma [ \tz{x}_k ] $ and 
$\StDvYt \equiv \sigma [ \tz{y}_k ] $, 
and that the errors on the $\tz{x}$ and $\tz{y}$ variables, for each measured
data point, are uncorrelated~\cite{bi:Macdonald,bi:PHBorcherdstAndCVShetht}.

The problem can be always reduced to an equivalent problem~\cite{bi:RBarlow}
with identical errors in both variables, by rescaling, via the standard
errors, to dimensionless variables, and multiplying, for the sake of generality,
by a common dimensionless factor, $\scaleFact$:
\begin{gather}
        \tz{x}_k \longrightarrow x_k \equiv \scaleFact \pton{\tz{x}_k / \StDvXt}
        \arrforw
        \sigma [ x_k ] = \scaleFact
        \spazio\spazio
        \tz{y}_k \longrightarrow y_k \equiv \scaleFact \pton{\tz{y}_k / \StDvYt}
        \arrforw
        \sigma [ y_k ] = \scaleFact
\punto
\end{gather}
The original variables, $ \tz{x}_k $ and $ \tz{y}_k $, will be called the
\textit{raw variables}, as opposed to the $x_k$ and $y_k$, the \textit{(re-scaled) variables}.
The above transformation will be always silently assumed in the rest of this paper.
In every physics problem this transformation is just a change of the units of measure of the two
variables, using $\StDvXt$ and $\StDvYt$ as the new units of measure,
leading to dimensionless variables.  One might choose $\scaleFact=1$,
but it is preferred to leave a generic $\scaleFact$ in order to have a better
understanding of the final formulas and deal with the case of a common but unknown error.

After the above transformation, the sum of distances between all the measured data
points and the straight line, 
(the error function) can be minimized, as a
purely geometrical problem. See~\cite{bi:RBarlow} for the precise
probabilistic/statistical discussion.

Note that there is no point in discussing any ambiguity of the best-fit line
under change of scale of the coordinates. In fact only when the errors in both
variables are equal one is allowed to minimize the distance between the measured
data points and the straight line, otherwise one needs to take into account the
error ellipse, see for instance~\cite{bi:RBarlow,bi:Lyons}.

The \mbox{\mbox{OLS-y:x}/\mbox{OLS-x:y}} fit can be recovered by letting
$\StDvXt\longrightarrow 0$/$\StDvYt\longrightarrow 0$ in the final results.

Finally, let us introduce the following notations for the variance and covariance
of the set of $N$ measured data points as whole, defined without the Bessel
correction factor $N/\pton{N-1}$, as it is useful in the formulation of the
LSFSL problem:
\begin{gather}
        \VarX \equiv \av{x^2} - {\av{x}}^2
        \spazio
        \VarY \equiv \av{y^2} - {\av{y}}^2   
        \spazio
        \CovXY  \equiv \av{xy} - \av{x}\av{y}
        \spazio
        \DD{V} \equiv \VarX - \VarY
\punto
\end{gather}

The above quantities, $\VarX$, $ \VarY$ and $\CovXY$, refer to the set
of measured data points as a whole, describing their spatial distribution in the $xy$
plane.  

The relations between variances and covariance in the raw and re-scaled coordinates, necessary for taking the
no-error limits, are obviously:
\begin{gather}
        \VarX \equiv \scaleFact^2 \afrac{\VarXt}{\StDvXt^2}
        \spazio
        \VarY \equiv \scaleFact^2 \afrac{\VarYt}{\StDvYt^2}
        \spazio
        \CovXY  \equiv \scaleFact^2 \afrac{\CovXYt}{\StDvXt\StDvYt}
\punto
\end{gather}

}

\nn{

Consider a set of $N$ measured data points, described by the $\tz{x}$ and $\tz{y}$ coordinates
of a suitable Cartesian Coordinate System: $\pgra{ \tz{x}_k ; \tz{y}_k }, ( k = 1, \ldots, N )$.
Assume that the measured data points follow the SWM, that is:
have equal standard errors, separately in both variables, 
$\StDvXt \equiv \sigma [ \tz{x}_k ] $ and 
$\StDvYt \equiv \sigma [ \tz{y}_k ] $, 
and that the errors on the $\tz{x}$ and $\tz{y}$ variables, for each measured data point, are uncorrelated:
\begin{gather}
        \label{eq:realValues}
        \pgra{ \tz{x}_k \pm \StDvXt ; \tz{y}_k \pm \StDvYt }
        \spazio
        (k = 1, \ldots, N)
        \spazio
        \rho_{\tz{x}\tz{y}} = 0
\punto
\end{gather}

It is assumed that each measured data point is a random sampling from a random
distribution associated to each true data point, $\pgra{ X_k ; Y_k }$, with true values lying on a unknown straight
line, to be determined:
\begin{gather}
        \label{eq:exactValues}
        \pgra{ X_k ; Y_k }
        \spazio
        A X_k + B Y_k + C = 0
        \spazio
        ( k = 1, \ldots, N )
        \\
        \pgra{ \tz{x}_k = X_k + \epsilon_{\tz{x}} ; \tz{y}_k = Y_k + \epsilon_{\tz{y}} }
        \spazio
        \text{with}
          \spazio
          \av{ \epsilon_{\tz{x}} } =0
          \spazio
          \sigma [ \epsilon_{\tz{x}} ] = \StDvXt
          \spazio
          \av{ \epsilon_{\tz{y}} } = 0
          \spazio
          \sigma [ \epsilon_{\tz{y}} ] = \StDvYt
\virgola
\end{gather}
where $ \epsilon_{\tz{x}} $ and $ \epsilon_{\tz{y}} $ are random variables,
often, but not always, Gaussian random variables.

The problem can be always reduced to an equivalent problem~\cite{bi:RBarlow}
with identical errors in both variables, by rescaling, via the standard
errors, to dimensionless variables, and multiplying, for the sake of generality,
by a common dimensionless factor, $\scaleFact$:
\begin{gather}
        \tz{x}_k \longrightarrow x_k \equiv \scaleFact \pton{\tz{x}_k / \StDvXt}
        \arrforw
        \sigma [ x_k ] = \scaleFact
        \spazio\spazio
        \tz{y}_k \longrightarrow y_k \equiv \scaleFact \pton{\tz{y}_k / \StDvYt}
        \arrforw
        \sigma [ y_k ] = \scaleFact
\punto
\end{gather}
The original variables, $ \tz{x}_k $ and $ \tz{y}_k $, will be called the
\textit{raw variables}, as opposed to the $x_k$ and $y_k$, the \textit{(re-scaled) variables}.
The above transformation will be always silently assumed in the rest of this paper.
In every physics problem this transformation is just a change of the units of measure of the two
variables, using $\StDvXt$ and $\StDvYt$ as the new units of measure,
leading to dimensionless variables.  One might choose $\scaleFact=1$,
but it is preferred to leave a generic $\scaleFact$ in order to have a better
understanding of the final formulas and deal with the case of a common but unknown error.

After the above transformation, the sum of distances between all the measured data
points and the straight line, 
(the error function) can be minimized, as a
purely geometrical problem. See~\cite{bi:RBarlow} for the precise
probabilistic/statistical discussion.

Note that there is no point in discussing any ambiguity of the best-fit line
under change of scale of the coordinates.  In fact only when the errors in both
variables are equal one is allowed to minimize the distance between the measured
data points and the straight line, otherwise one needs to take into account the
error ellipse, see for instance~\cite{bi:RBarlow,bi:Lyons}.

The \mbox{\mbox{OLS-y:x}/\mbox{OLS-x:y}} fit can be recovered by letting
$\StDvXt\longrightarrow 0$/$\StDvYt\longrightarrow 0$ in the final results.

Finally, let us introduce the following notations for the variance and covariance
of the set of $N$ measured data points as whole, defined without the Bessel
correction factor $N/\pton{N-1}$, as it is useful in the formulation of the
LSFSL problem:
\begin{gather}
        \VarX \equiv \av{x^2} - {\av{x}}^2
        \spazio
        \VarY \equiv \av{y^2} - {\av{y}}^2   
        \spazio
        \CovXY  \equiv \av{xy} - \av{x}\av{y}
        \spazio
        \DD{V} \equiv \VarX - \VarY
\punto
\end{gather}
The above quantities, $\VarX$, $ \VarY$ and $\CovXY$, refer to the set
of measured data points as a whole, describing their spatial distribution in the $xy$
plane: they are obviously distinct from the spread of one single measurement around its true
value.  

In order to avoid any confusion, the symbols $\VarX$, $ \VarY$ and $\CovXY$, are
used in this paper to refer to the set of measured data points as a whole, while the
symbols $\StDvX^2$, $\StDvY^2$ and $\rho_{xy} = 0$ are used to refer to the
variances and correlation coefficient of the two-dimensional random variable associated to each
specific measured data point.

The relations between variances and covariance in the raw and re-scaled coordinates, necessary for taking the
no-error limits, are obviously:
\begin{gather}
        \VarX \equiv \scaleFact^2 \afrac{\VarXt}{\StDvXt^2}
        \spazio
        \VarY \equiv \scaleFact^2 \afrac{\VarYt}{\StDvYt^2}
        \spazio
        \CovXY  \equiv \scaleFact^2 \afrac{\CovXYt}{\StDvXt\StDvYt}
\punto
\end{gather}
}

\section{The slope/intercept parametrization}
\label{pa:SlopeIntercept}

The least squares fit to the straight lines
\begin{gather}\label{eq:SlopeIntercept}
        y = p_y x + q_y
        \spazio
        p_y = \tan\pqua{\thx}
        \spazio\text{or}\spazio
        x = p_x y + q_x
        \spazio
        p_x = \tan\pqua{\thy}
\virgola
\end{gather}  
in terms of the slopes, $p_y$/$p_x$, intercepts, $q_y$/$q_x$, and angles
$\thx$/$\thy$, with respect to the $x$/$y$ axes, is appropriate in many physics
problems, for instance when fitting a position as a function of time, so that
the slope (velocity) can be zero but not infinity.

Some of the most well-known
textbooks~\cite{bi:Bevington,bi:Cowan,bi:Taylor,bi:Zech,bi:Lyons} use
this parametrization to describe straight lines and present the so-called
\textit{effective variance method}~\cite{bi:Orear}, to deal with measured data points
having significant errors in both variables.  
One exception is textbook~\cite{bi:RBarlow}, which presents the exact
solution within the SWM.

The dimensionless error function to minimize, the sum of distances between all the measured data
points and the straight line, is
\begin{gather}\label{eq:ChiSqSlopeIntercept}
        \chiSq_y[p_y,q_y] \equiv \afrac{1}{\scaleFact^2}\sum_{k=1}^N \afrac{\pton{ p_y x_k + q_y - y_k }^2}{ 1 + p_y^2 }
        \spazio\text{or}\spazio
        \chiSq_x[p_x,q_x] \equiv \afrac{1}{\scaleFact^2}\sum_{k=1}^N \afrac{\pton{ p_x y_k + q_x - x_k }^2}{ 1 + p_x^2 }
\virgola
\end{gather} 
which will be called $\chiSq$, regardless of its statistical properties.

Minimization of the error functions leads, after some mathematics, to
the expressions for the slopes and intercepts
which can be found in some of the
literature~\cite{bi:Ross,bi:RBarlow,bi:Macdonald}.
The solution, in both the $y$ versus $x$ and the $x$ versus $y$ representations, is:
\begin{gather}
        \text{for $ \CovXY \neq 0 $}
        \spazio
        \text{with $ \alpha \equiv \sign\pqua{\CovXY} \equiv \afrac{\CovXY}{\abs{\CovXY}}$}
        \\
        \label{eq:SlopeInterceptYSol}
        A_y \equiv \afrac{ \VarY - \VarX }{ 2 \CovXY }
        \spazio
        p_y = 
        A_y + \alpha \sqrt{1+A_y^2} =
        \afrac{ ( \VarY - \VarX ) + \sqrt{ 4 \CovXY^2 + \pton{\DD{V}}^2 } }{ 2 \CovXY }
        \spazio
        q_y = \av{y} - p_y \av{x}
        \\
        \label{eq:SlopeInterceptXSol}
        A_x \equiv \afrac{ \VarX - \VarY }{ 2 \CovXY }
        \spazio
        p_x = 
        A_x + \alpha \sqrt{1+A_x^2} =
        \afrac{ ( \VarX - \VarY ) + \sqrt{ 4 \CovXY^2 + \pton{\DD{V}}^2 } }{ 2 \CovXY }
        \spazio
        q_x = \av{x} - p_x \av{y} 
        \\
        \label{eq:AngsDefs}
        \text{with $ p_y p_x = 1 $}
        \spazio
        \text{and $ \thx + \thy = \pi/2 $}
\punto
\end{gather}

Note that the sign in front of the square-root is unambiguously determined by $\alpha$.

The above relation $ p_y p_x = 1 $ is between the two different representations of the same
best-fit straight line. It should not be confused with the relation between 
the $ p_y $ and $ p_x $ slopes as determined by the \mbox{\mbox{OLS-y:x}/\mbox{OLS-x:y}} fit,
giving the correlation coefficient, $r$, of the set of data-points as whole (see
for instance~\cite{bi:Young}): $p_y p_x = r^2$.

Whenever $ \CovXY = 0 $ the fitted line is parallel to either the $x$ or $y$ axis.
The sign of the slopes is always the same as the sign of the $ \CovXY $.
The straight line perpendicular to the minimum-distance straight line maximizes the error
function; its slope can be found by putting, in the above formulas, a minus
sign, instead of a plus sign, in front of the square root. 

A simple trigonometric transformation allows to transform
equations~\ref{eq:SlopeInterceptYSol} and~\ref{eq:SlopeInterceptXSol}
into an even simpler equation for the angles $\thx$ and $\thy$:
\begin{gather}\label{eq:SlopeInterceptSolTheta}
         \tan\pqua{2\thx} = -\afrac{1}{A_y}
         \spazio
         \text{for $A_y \neq 0$}
         \spazio
         \text{or}
         \spazio
         \tan\pqua{2\thy} = -\afrac{1}{A_x}
         \spazio
         \text{for $A_x \neq 0$}
\punto
\end{gather}

It is useful to re-write the above
equations~\ref{eq:SlopeInterceptYSol}
and~\ref{eq:SlopeInterceptXSol} as a function of the raw variables,
which also allows to take the limit of the \mbox{\mbox{OLS-y:x}/\mbox{OLS-x:y}} fit.
One finds~\cite{bi:RBarlow}:
\begin{gather}
        \label{eq:SlopeInterceptYSolSlopeOrig}
        p_{\tz{y}} \equiv \tan\pqua{\thtx} =
        \afrac{\StDvYt}{\StDvXt} \pton{ A_y + \alpha \sqrt{1+A_y^2} } 
        \spazio
        A_y \equiv \afrac{ \StDvXt^2 \VarYt - \StDvYt^2 \VarXt }{ 2 \StDvXt\StDvYt\CovXYt }
\virgola
        \\
        \label{eq:SlopeInterceptXSolSlopeOrig}
        p_{\tz{x}} \equiv \tan\pqua{\thty} =
        \afrac{\StDvXt}{\StDvYt} \pton{ A_x + \alpha \sqrt{1+A_x^2} } 
        \spazio
        A_x \equiv \afrac{ \StDvYt^2 \VarXt - \StDvXt^2 \VarYt }{ 2 \StDvXt\StDvYt\CovXYt }
\punto
\end{gather}

The standard errors on the slope and intercept will be derived in
section~\ref{pa:AngleAndSignedDistanceErrorsOnPQ}.

\section{The angle/signed-distance parametrization}
\label{pa:AngleAndSignedDistance}

\subsection{Parametrization of a straight line in the plane}

In many problems the $x$ and $y$ coordinates are equivalent, such as, for
instance, in straight line fitting of pixelized images with rectangular/square pixels.
Especially in these case a parametrization of the straight line different from the most common
slope/intercept parametrization may be useful, as the direction of the straight
line is better identified by the angle with respect to one specific direction in
the plane, for instance the $x$ axis, than by the slope.  In fact both the slope
and intercept may tend to infinity for straight lines passing near the origin
and nearly parallel to the $y$ or $x$
axis (depending on the representation used from
equation~\ref{eq:SlopeIntercept}).

\pp{

The geometrical aspects of the angle/signed-distance parametrization of a straight line in the plane
are described in this section.

Any straight line in the $xy$ plane
can be described by two real parameters identifying in a unique way all straight
lines in the plane, such as:
\begin{itemize}
\item
the angle, $ \theta $, between the straight line and the $x$ axis:
\begin{gather}\label{eq:AngleAndSignedDistanceThe}
        -\pi/2 < \theta \leq +\pi/2 \arrforw \cos\pqua{\theta} \geq 0
        \spazio\text{the range of $\theta$ is just one possible conventional choice}
\puntoevirgola
\end{gather}
\item
the signed-distance $ c $, given by the $z$
component of the vector product between the unit direction vector of the
straight line, $\ve{n}$, and any point on the straight line, $\ve{R}$, whose absolute value
gives the distance between the straight line and the origin:
\begin{gather}\label{eq:AngleAndSignedDistanceCpx}
        -\infty < c < +\infty 
        \spazio
        c \equiv \ve{e}_z \cdot \pton{ \ve{n} \times \ve{R} }
        \spazio\text{for any point $ \ve{R} $ belonging to the straight line}
\punto
\end{gather}
\end{itemize}

The geometry is represented in Figure~\ref{fi:graph}.
\begin{figure}[htb]
  \centering
  \includegraphics[width=0.60\textwidth]{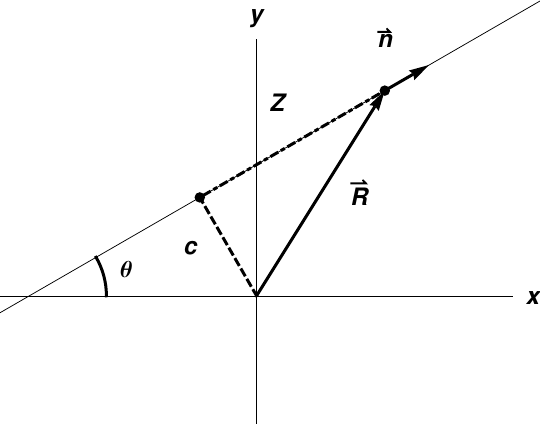}            
  \caption{The angle/signed-distance parametrization.}
\label{fi:graph}
\end{figure}

Converting into the Cartesian representation one finds:
\begin{gather}\label{eq:AngleAndSignedDistanceParameterization}
        x \sin\pqua{\theta} - y \cos\pqua{\theta} + c = 0
\punto
\end{gather}

}

\nn{

The details of the geometrical aspects of the
parametrization of a straight line in the plane
which are useful for the understanding of the results are described in this section.

Any straight line in the $xy$ plane passing by the point $ \ve{R} \equiv \pgra{ X ; Y } $
can be represented in terms of the direction unit vector, 
$ \ve{n} \equiv \pgra{ \cos\pqua{\theta} ; \sin\pqua{\theta} } $, by the parametric equation
\begin{gather}\label{eq:AngleAndSignedDistanceParametricParameterization}
        \ve{r}[t] = \ve{R} + \ve{n} t 
        \spazio
        \ve{n} \equiv \pgra{ \cos\pqua{\theta} ; \sin\pqua{\theta} }
        \spazio
        \ve{R} \equiv \pgra{ X ; Y }
        \spazio
        \begin{cases}
                x[t] = X + t \cos\pqua{\theta}
                \\
                y[t] = Y + t \sin\pqua{\theta}
        \end{cases}
\punto
\end{gather}

Two real parameters identifying in a unique way all straight lines in the plane
can be chosen as:
\begin{itemize}
\item
the angle, $ \theta $, between the straight line and the
$x$ axis:
\begin{gather}\label{eq:AngleAndSignedDistanceThe}
        -\pi/2 < \theta \leq +\pi/2 \arrforw \cos\pqua{\theta} \geq 0
        \spazio\text{the range of $\theta$ is just one possible conventional choice}
\puntoevirgola
\end{gather}
\item
the signed-distance $ c $, given by the $z$
component of the vector product between $\ve{n}$ and $\ve{r}$, whose absolute value
gives the distance between the straight line and the origin:
\begin{gather}\label{eq:AngleAndSignedDistanceCpx}
        -\infty < c < +\infty 
        \spazio
        c \equiv \ve{e}_z \cdot \pton{ \ve{n} \times \ve{R} }
        \spazio\text{for any point $ \ve{R} $ belonging to the straight line}
\puntoevirgola
\end{gather}
in fact the magnitude and sign of $c$ is also
given by the vector product between $ \ve{n} $ and the vector
position of any point on the line, $\ve{R}$;
the sign of $c$ is always the same as the sign of $q_y$.
\end{itemize}

The geometry is represented in Figure~\ref{fi:graph}.
\begin{figure}[htb]
  \centering
  \includegraphics[width=0.40\textwidth]{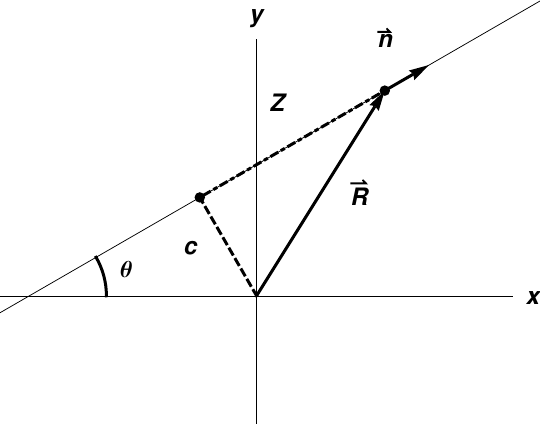}            
  \caption{The angle/signed-distance parametrization.}
\label{fi:graph}
\end{figure}

Converting into the Cartesian representation one finds:
\begin{gather}\label{eq:AngleAndSignedDistanceParameterization}
        x \sin\pqua{\theta} - y \cos\pqua{\theta} + c = 0
        \\
        c = - X \sin\pqua{\theta} + Y \cos\pqua{\theta}
        \spazio\text{for any point $ \ve{R} = \pgra{X;Y} $ belonging to the straight line}
        \\
        \pton{ x - X } \sin\pqua{\theta} - \pton{ y - Y } \cos\pqua{\theta} = 0
\punto
\end{gather}

}

Of course, other parameterizations are possible, but the above parametrization
has the virtue of describing bi-univocally all straight lines in the plane in terms of
two real non-singular parameters with a clear and direct geometrical interpretation.
Moreover the two parameters $\theta$ and $c$ have simple transformation properties
under translation and rotations in the $xy$ plane, at variance with the slope and intercept.

\subsection{Determination of the best-fit line}
\label{pa:Sol}

The error function to minimize using the
angle/signed-distance parametrization is: 
\begin{gather}\label{eq:ChiSquare}
        \chiSq[\theta,c] \equiv 
        \afrac{1}{\scaleFact^2}\sum_{k=1}^N \pton{ x_k \sin\pqua{\theta} - y_k \cos\pqua{\theta} + c }^2
\virgola
\end{gather}
which will be called $\chiSq$, regardless of its statistical properties.

Note that the problem is a non-linear least squares problem, so that the nice
general properties of linear least squares problems (see for
instance~\cite{bi:James}) are not guaranteed.
In particular the exact contour of the confidence region in the parameter space is
not elliptical~\cite{bi:Cowan}.

\nn{
All the details of the minimization of the error function in equation~\ref{eq:ChiSquare} are
summarized in appendix~\ref{app:ThetaAndsignedDistanceResults},
where all the following results are derived.
}

\pp{
All the details of the minimization of the error function in equation~\ref{eq:ChiSquare} are
summarized in~\cite{bi:nota},
where all the following results are derived.
}

Equating to zero the derivative with respect to $c$ of equation~\ref{eq:ChiSquare}, in order to look for
stationary points of the error function, immediately provides the condition that any
straight line leading to a stationary point of the error function passes by the centroid of the data
points, $\ve{C} \equiv \pgra{ \av{x} ; \av{y} }$, (as it is the case for the 
\mbox{\mbox{OLS-y:x}/\mbox{OLS-x:y}} fit): 
\begin{gather}\label{eq:SignedDistanceFormula}
        \pderiv{\chiSq[\theta,c]}{c} = 0
        \longrightarrow
        \est{c} = - \av{x} \sin\pqua{\est{\theta}} + \av{y} \cos\pqua{\est{\theta}} 
\punto
\end{gather}  

The error function in equation~\ref{eq:ChiSquare} becomes a function of $\theta$ only, after using
equation~\ref{eq:SignedDistanceFormula} to replace the signed-distance.
It can be written as:
\begin{gather}\label{eq:ChiSquareMinThetaOnly}
        \afrac{ \scaleFact^2 \chiSq[\theta] }{N} \equiv 
        \afrac{1}{N} \sum_{k=1}^N 
                \pton{ \pton{ x_k - \av{x} } \sin\pqua{\theta} - \pton{ y_k - \av{y} } \cos\pqua{\theta} }^2 =
        \VarX \sin^2\pqua{\theta} + \VarY \cos^2\pqua{\theta} - 2 \CovXY \sin\pqua{\theta}\cos\pqua{\theta}
\punto
\end{gather}

The values of $\theta$ giving the stationary points are 
\nn{(see appendix~\ref{app:ThetaAndsignedDistanceResults} for the derivation of the results)}:
\pp{(see~\cite{bi:nota} for the derivation of the results)}:
\begin{gather}
         \DD{V} = 0
         \arrforw
         \cos\pqua{ 2\theta }=0
         \arrforw
         \est{\theta} = \afrac{\pi}{4} + q\afrac{\pi}{2}
         \spazio q \in \mathbb{Z}
         \\
         \DD{V} \neq 0
         \arrforw
         \tan\pqua{ 2\theta } = \afrac{2 \CovXY }{ \DD{V} }
         \arrforw
         \est{\theta} = \afrac{1}{2}\arctan\pqua{\afrac{2 \CovXY }{ \DD{V} }} + q\afrac{\pi}{2}
         \spazio q \in \mathbb{Z}
         \\
         \notag
         \text{minimum: choose $\est{\theta}$ such that $|\est{\theta}|\lt\pi/2$ with the same sign as $\CovXY$}
\punto
\end{gather}

In the often encountered case that the common error on the data points, $\scaleFact$, is
unknown, it can be estimated from the data.
It can be shown~\cite{bi:James} that, in the linear least squares case, the residual sum of squares
at the minimum, divided by the number of degrees of freedom, $N-2$ (as two parameters are determined by the
minimization), is an unbiased estimate of the unknown error $\scaleFact$:
\begin{equation}
        \est{\scaleFact^2} = S^2 \equiv \afrac{\chiSq_{\textrm{min}}}{N-2}
\punto
\end{equation}

\subsection{Formulas for the standard errors}

The simple exact analytic formulas for the standard errors and covariance of the
fit parameters are presented and discussed in this section.  Exact means that
they can be derived without any approximation from the standard formula for the
propagation of errors~\cite{bi:HughesHase,bi:RBarlow,bi:Cowan,bi:Zech,bi:Lyons,bi:James},
which is derived using the first-order truncated Taylor series expansion of the
function and taking expectations values.  Approximate formulas were published
in~\cite{bi:PHBorcherdstAndCVShetht}.

A long and tedious, but straightforward, direct calculation, applying the standard
formula for the propagation of errors to equation~\ref{eq:SlopeInterceptSolTheta}, leads to the following 
results for the standard errors 
\nn{(see the appendix~\ref{app:ThetaAndsignedDistanceResults} for details of the calculation)}:
\pp{(see~\cite{bi:nota} for details of the calculation)}:
\begin{gather}
        \ve{n} \equiv \pgra{ \cos\pqua{\theta} ; \sin\pqua{\theta} } 
        \spazio
        \ve{C} \equiv \pgra{ \av{x} ; \av{y} } 
        \spazio
        Z \equiv \av{x} \cos\pqua{\est{\theta}} + \av{y} \sin\pqua{\est{\theta}}
        \equiv \ve{n} \cdot \ve{C}
        \\
        \label{eq:dtheta}
        \variance\pqua{\est{\theta}} \equiv \pton{\ddd{\theta}}^2 = 
        \afrac{\scaleFact^2}{N} \afrac{ \VarX + \VarY }{ \pton{\DD{V}}^2 + 4 \CovXY^2 }
        \\
        \label{eq:dc}
        \variance\pqua{\est{c}} \equiv \pton{\ddd{c}}^2 = 
        \afrac{\scaleFact^2}{N} + \pton{Z \ddd{\theta}}^2
        \\
        \label{eq:dctheta}
        \covariance\pqua{\est{\theta},\est{c}} = 
        -Z \pton{\ddd{\theta}}^2
\punto
\end{gather}

Equations~\ref{eq:dtheta},~\ref{eq:dc} and~\ref{eq:dctheta} generalize the
well-known results for the \mbox{\mbox{OLS-y:x}/\mbox{OLS-x:y}} fit.  They can all be expressed in
terms of the standard error on the angle plus the geometrical term $Z$.

$Z$ is interpreted as the signed-projection of the position
vector of the centroid, $ \ve{C}$, onto the direction unit vector, $\ve{n}$, of the
best-fit line. The absolute value of $Z$ is the distance between the centroid
and the straight line perpendicular to the best-fit line and passing by the
origin.

All the above standard errors, equations~\ref{eq:dtheta},~\ref{eq:dc}
and~\ref{eq:dctheta}, must be invariant under rotations in the $xy$ plane as $c$
is invariant while $\theta$ is just shifted under rotations.  However this
property is not obvious from equations~\ref{eq:dtheta},~\ref{eq:dc}
and~\ref{eq:dctheta}; it is demonstrated in section~\ref{pa:PCA}.  

The
standard error in equation~\ref{eq:dtheta} must be also invariant under
translations in the $xy$ plane, because $\theta$ is invariant under
translations, while $c$ is not.  This property is obvious from
equations~\ref{eq:dtheta},~\ref{eq:dc} and~\ref{eq:dctheta}, as only $Z$ 
is not invariant under translations in
equations~\ref{eq:dtheta},~\ref{eq:dc} and~\ref{eq:dctheta}.

Equation~\ref{eq:dc} shows that the error on the signed-distance $c$ is
minimized if the origin of the Cartesian Coordinate System is set at the
centroid of the measured data points, so that $Z=0$ (the same property applies
in \mbox{\mbox{OLS-y:x}/\mbox{OLS-x:y}} fit). In this case the error on
the signed-distance $c$ is just the expected $ \scaleFact/\sqrt{N} $, that is
the error on the determination of the mean values $\av{x}$ and $\av{y}$,
according to equation~\ref{eq:SignedDistanceFormula}.  Moreover if $Z=0$ the
estimates of $\theta$ and $c$ are uncorrelated.  In the general case
equation~\ref{eq:dc} is the sum in quadrature of a term coming from the
uncertainty in the position of the centroid, the first one, plus a term coming
from the uncertainty in the angle, amplified by the distance $|Z|$.

It should be emphasized, as pointed out in some of the literature, that the formulas for the standard errors
should be evaluated, in principle, via the true data points and not via the measured data
points, as the
standard error expressions are derived from a Taylor series expansion about the
true data points.
In practice the measured data points are normally
used to evaluate the standard error expressions. The accuracy of this approximation
is studied in~\cite{bi:nota}, via Monte Carlo simulations.
In principle, after the fit, one might want to correct the measured data points to estimate the
true data points values, improving the calculation of the standard errors.
However it is shown in section~\cite{bi:nota} that this is normally not necessary. 

\subsection{Bias}

Since the least squares problem of Eq.~\ref{eq:ChiSquare}
is a non-linear one, a possible bias of the estimates
must be studied.

The bias of the estimated parameter $\est{\theta}$ is invariant under translations and rotations
of the Cartesian coordinate system, because $\theta$ itself is
invariant under translations and it is just shifted by a constant under rotations.  It
can be shown that $\est{\theta}$ is un-biased as follows.  Imagine
setting the origin of the Cartesian coordinate system at the unknown position of
the centroid of the true data points, lying on the true straight line, with one
axis along the true straight line, so that $\theta=0$.  
By symmetry, the resulting probability
distributions of all the measured data points are symmetrical about the unknown
true straight line.  Therefore, for any configuration of measured data points,
there is another configuration having all the measured data points located
symmetrically with respect to the true straight line; these two symmetrical configurations,
having equal probability, would give two opposite values for the estimate of
$\theta$.  Therefore the distribution of $\est{\theta}$ is symmetrical about the
true value, that is zero, so that the estimation of $\theta$ is un-biased.

A similar reasoning leads to a determination of the bias of the estimator $\est{c}$ of the
signed-distance.  
Although $c$ is invariant under rotations of the Cartesian
coordinate system, its change under translations depends on the angle
$\theta$.  
Imagine rotating the Cartesian coordinate system in such a way that the true
straight line has $\theta=0$.  
By symmetry, the resulting probability
distributions of all the measured data points are symmetrical about the unknown true
straight line.  
Therefore, for any configuration of measured data points, there
is one configuration having all the measured data points located symmetrically
with respect to the true straight line.
These two symmetrical configurations, having equal probability, would give: two
opposite values for the estimate of $\theta$; the same value of the $x$
coordinate of the centroid of the measured data points; two different values,
which average to $c$, for the $y$ coordinate of the centroid of the measured data
points.
Therefore, using a simple geometrical
construction, one can show that the average value of the two values of
$\est{c}$ determined from these two symmetrical configurations of measured data points 
is $ c \cos\theta $, so that the bias is:
\begin{equation}\label{eq:CBias}
        \av{ \est{c} - c } =  c \av{\cos\pton{\est{\theta}-\theta}} - c
\punto
\end{equation}
Moreover,
\begin{equation}
        \abs{ \av{ \est{c} - c } } \leq  2c 
        \virgola
        \spazio
        \abs{ \av{ \est{c} - c } } \leq  \afrac{c}{2} \variance\pton{ \est{\theta} - \theta } + \mathcal{O}\pton{ \est{\theta} - \theta }^4
\punto
\end{equation}
Using the same geometrical construction, the same result can be obtained algebraically from
Eq.~(\ref{eq:SignedDistanceFormula}), taking into account that one has: 
$ \est{\theta}_1 + \est{\theta}_2 = 0 $, 
$ \av{x}_1 = \av{x}_2 $ and 
$ \av{y}_1 + \av{y}_2 = 2c $, for the two
symmetrical configurations defined above.

\subsection{The covariance matrix and Principal Components Analysis}
\label{pa:PCA}

Principal Components Analysis (PCA) provides a better insight into the results
of the LSFSL-SWM and help to derive some results in a simple way.  It relies on
orthogonal transformations of random variables~\cite{bi:Cowan}. 
PCA and its
relation with LSFSL-SWM, discovered by
K. Pearson~\cite{bi:Pearson}, are extensively discussed in~\cite{bi:PCA}.

In the rotated Cartesian Coordinate System 
$ \pton{ x ; y } \rightarrow \pton{ \xi ; \eta } $, such that $\theta$ is the
angle made by the $\xi$ axis with the $x$ axis, one has:
\begin{gather}
        \xi   = + x \cos\pqua{\theta} + y \sin\pqua{\theta}
        \spazio
        \eta  = - x \sin\pqua{\theta} + y \cos\pqua{\theta}
\virgola
\end{gather}
and the variances of the set of data points are readily calculated:
\begin{gather}\label{eq:MaxAndMinVariance}
\begin{cases}
        \VarCsi =
        \VarX \cos^2\pqua{\theta} + \VarY \sin^2\pqua{\theta} + 2 \CovXY \sin\pqua{\theta}\cos\pqua{\theta}
\\
        \VarEta = 
        \VarX \sin^2\pqua{\theta} + \VarY \cos^2\pqua{\theta} - 2 \CovXY \sin\pqua{\theta}\cos\pqua{\theta}=
        \VarX + \VarY - \VarCsi
\punto
\end{cases}
\end{gather}

Therefore requiring that the variance, $\VarEta$, of the set of measured data
points as a whole is minimized, leads to exactly the same
results as minimizing equation~\ref{eq:ChiSquareMinThetaOnly}. In fact
the LSFSL-SWM minimizes the sum of the distances between the measured data points and the
straight line at angle $\theta$ from the $x$ axis, namely the $\xi$ axis.

Equations~\ref{eq:MaxAndMinVariance} show the well-known
result that the sum of the variances is invariant under rotations, as it is the
trace of the covariance matrix. Therefore minimizing $\VarEta$ is equivalent to
maximizing $\VarCsi$.
Some standard linear algebra, see for instance~\cite{bi:PCA},
shows that the direction $\xi$ maximizing the variance $ \VarCsi $ in
equation~\ref{eq:MaxAndMinVariance} can be found by diagonalizing the covariance matrix:
\begin{gather}
        \label{eq:CMatrix}
        \ve{C} \equiv
        \begin{pmatrix}
          \VarX         &       \CovXY
          \\
          \CovXY        &       \VarY
        \end{pmatrix}
        \longrightarrow
        \ve{C}^\prime \equiv
        \begin{pmatrix}
          \VarCsi       &       \CovXiEta=0
          \\
          \CovXiEta=0   &       \VarEta
        \end{pmatrix}
\punto
\end{gather}

Therefore diagonalizing the covariance matrix leads to maximize the variance
along the $\xi$ axis and minimize the variance along the $\eta$ axis: the
largest eigenvalue, $ \lambda_{+} $, is the variance along the $\xi$ axis and the smallest
eigenvalue, $\lambda_{-}$, is the variance along the $\eta$ axis.
The two orthonormal eigenvectors give:
the direction maximizing the variance, namely the $\xi$ axis, (the eigenvector
corresponding to the largest eigenvalue, $\pgra{ \cos\pqua{\theta} ; \sin\pqua{\theta} }$);
the direction minimizing the variance, namely the $\eta$ axis, (the eigenvector
corresponding to the smallest eigenvalue), corresponding to minimize 
the error function in equation~\ref{eq:ChiSquare}.

The following results can be obtained:
\nn{
\begin{gather}
        \lambda_{\pm} = \afrac{1}{2} \pton{ \pton{ \VarX + \VarY } \pm \sqrt{ \pton{\DD{V}}^2 + 4 \CovXY^2 } }
        \label{eq:lambda}
        \\
        0 \leq \lambda_{-} \leq \lambda_{+}
        \spazio
        \lambda_{+} + \lambda_{-} = \VarX + \VarY
        \spazio
        \lambda_{+} - \lambda_{-} = \sqrt{ \pton{\DD{V}}^2 + 4 \CovXY^2 }
        \label{eq:lambdaProp}
\punto
\end{gather}
}
\pp{
\begin{gather}
        \lambda_{\pm} = \afrac{1}{2} \pton{ \pton{ \VarX + \VarY } \pm \sqrt{ \pton{\DD{V}}^2 + 4 \CovXY^2 } }
        \label{eq:lambda}
\punto
\end{gather}
}

The case of perfect linear correlation corresponds to $ \CovXY^2 = \VarX \VarY $,
that is zero determinant of the covariance matrix.

\subsection{Some properties of the standard errors}

The discussion in section~\ref{pa:PCA} gives some insight into the understanding of the properties
of the standard errors in
equations~\ref{eq:dtheta},~\ref{eq:dc} and~\ref{eq:dctheta}.

The expressions for the eigenvalues of the covariance matrix,
equations~\ref{eq:lambda}, show that the standard errors in
equations~\ref{eq:dtheta},~\ref{eq:dc} and~\ref{eq:dctheta} are invariant under
rotation in the $xy$ plane, as they can be written in terms of the eigenvalues of
the covariance matrix and the signed-distance $Z$.

In the rotated Cartesian Coordinate System $\pton{ \xi ; \eta }$, 
one has:
\begin{gather}
        \VarCsi \equiv \lambda_{+}
\spazio
        \VarEta \equiv \lambda_{-}
\spazio
        \pton{\ddd{\theta}}^2 = \afrac{\scaleFact^2}{N} 
        \afrac{ \VarCsi + \VarEta }{ \pton{ \VarCsi - \VarEta }^2 }
\punto
\end{gather}

In the often encountered case that the measured data points are highly linearly correlated, the
covariance can be approximately written in terms of the variances, 
$\CovXY^2 \simeq \VarX \VarY$, and equation~\ref{eq:dtheta} simplifies to:
\begin{gather}
        \label{eq:ErrorThetaGoodCorrelation}
        \pton{\ddd{\theta}}^2 \simeq
        \afrac{\scaleFact^2}{N} \afrac{1}{ \VarX + \VarY } \simeq
        \afrac{\scaleFact^2}{N} \afrac{1}{ \VarCsi } 
\punto
\end{gather}

The above result for highly linearly correlated measured data points has a very simple
interpretation. 
It shows that the standard error on the
angle can be expressed as the ratio between the single data point standard error,
$\scaleFact$, and the square-root of the variance along the best-fit straight line 
of the set of measured data points as a
whole, $\sqrt{\VarCsi}$, divided by the square-root
of the number of measured data points. Therefore the standard error on the
angle can be interpreted as the 
transverse size of the single measured data point, $\scaleFact$,
divided by an 
effective length of the measured data points set, 
given by $\sqrt{\VarCsi}$, similarly to the definition of the
radian, all divided by the square-root
of the number of measured data points.

This result can be used to easily estimate the error expected from the
LSFSL-SWM, for highly linearly correlated data.
Moreover it makes quantitative the naive expectation that the data
points at the two extremes of the straight line have more importance for the
fit, as they increase the variance $\VarCsi$ decreasing the error on the angle.
In fact: $ \ddd{\VarCsi} = 2 \pton{\xi_j - \av{\xi}} \ddd{\xi_j} / N $.

\subsection{Standard errors on the slope and intercept}
\label{pa:AngleAndSignedDistanceErrorsOnPQ}

Using the results derived in this section it is possible to find the formulas
for the standard errors on the slope and intercept:
\begin{gather}
        \label{eq:ErrorSlope2}
        \pton{\ddd{p}}^2 = \pton{ 1 + p^2 }^2 \pton{\ddd{\theta}}^2
        \\
        \label{eq:ErrorIntercept2}
        \pton{\ddd{q}}^2 = 
        \pton{ 1 + p^2 } \pton{ \afrac{\scaleFact^2}{N} + \pton{\ddd{\theta}}^2 \pton{Z - c p }^2 } =
        \pton{ 1 + p^2 } \pton{ \afrac{\scaleFact^2}{N} + \pton{\ddd{\theta}}^2 \av{x}^2 \pton{ 1 + p^2 } }
        \\
        \label{eq:CovPQ2}
        \covariance\pqua{p,q} = 
        - \pton{ 1 + p^2 }^{3/2} \pton{Z - c p } \pton{\ddd{\theta}}^2 =
        - \pton{ 1 + p^2 }^2 \av{x} \pton{\ddd{\theta}}^2
\punto
\end{gather}
\nn{
See appendix~\ref{app:ErrorsSlopeIntercept} for the details.
}
\pp{
See~\cite{bi:nota} for the details.
}

\subsection{Comparison with the \texorpdfstring{\mbox{\mbox{OLS-y:x}/\mbox{OLS-x:y}}}{.....} fit}


The invariance under rotation of the error function (as it is a sum of
distances) can be used to compare the results 
with the \mbox{\mbox{OLS-y:x}/\mbox{OLS-x:y}} fit.

The error functions in equation~\ref{eq:ChiSqSlopeIntercept} clearly show that
the results of the fit
will tend to the results for the \mbox{\mbox{OLS-y:x}/\mbox{OLS-x:y}} 
fit whenever $ p_y \simeq 0 $/$ p_x \simeq 0 $, as in this case, the distance is measured in a direction
which is both perpendicular to the straight line and parallel to the $y$/$x$
axis.

In fact, after determining the best-fit straight line one can apply a roto-translation to bring 
the origin of a new Cartesian Coordinate System, $\pgra{ \xi ; \eta }$,
coincident with the centroid of the measured data points, and the $\xi$ axis
along the best-fit
line. In this new Cartesian Coordinate System the error
function in equation~\ref{eq:ChiSquare} is exactly the same as it would be for the 
\mbox{OLS-$\eta$:$\xi$} fit. 
The covariance of the set of measured data points, $\CovXiEta$,
is zero and the best-fit slope and intercept are both
zero, in both cases by construction of the Cartesian Coordinate System $\pgra{ \xi ; \eta }$.


\nn{
A more detailed discussion can be found in appendix~\ref{app:LimOLS}.
}
\pp{
It can be shown, see~\cite{bi:nota}, that the formulas for the errors in the
slope/intercept parameterization,
equations~\ref{eq:ErrorSlope2},~\ref{eq:ErrorIntercept2} and~\ref{eq:CovPQ2},
tend to the well-known formulas in the \mbox{\mbox{OLS-y:x}/\mbox{OLS-x:y}} fit.
}

\subsection{OLS - negligible error in one of the variables}

In the case of the highly linearly correlated measured data points of
equation~\ref{eq:ErrorThetaGoodCorrelation}, it is easy to find a simple
criterion to assess whenever the errors in one of the two variables can
be neglected, the OLS fit.

After re-writing equation~\ref{eq:ErrorThetaGoodCorrelation} in terms of the raw
variables one finds:
\begin{gather}\label{eq:Criterium1}
        \pton{\ddd{\theta}}^2 = 
        \afrac{\scaleFact^2}{N} \afrac{1}{ \VarX + \VarY } =
        \afrac{1}{N} \afrac{1}{ \afrac{\VarXt}{\StDvXt^2} + \afrac{\VarYt}{\StDvYt^2} } 
\punto
\end{gather}

Note, however, that equation~\ref{eq:Criterium1} does not give the
limiting standard error on the raw angle $\tz{\theta}$, which is the interesting
quantity, but the one on the re-scaled angle $\theta$. 

The complete analysis, 
\nn{in the appendix~\ref{app:LimOLS},}
\pp{see~\cite{bi:nota},}
shows that the error on
$x$/$y$ is negligible depending on the relative values of 
$ \VarXt / \StDvXt^2 $ versus $ \VarYt / \StDvYt^2 $: 
\begin{gather}\label{eq:Criterium2}
  \begin{cases}
         \afrac{\VarXt}{\StDvXt^2} \gg \afrac{\VarYt}{\StDvYt^2}
         \arrforw
         \text{\mbox{OLS-y:x}}
         \spazio
         \ddd{ p_{\tz{y}} } \longrightarrow \afrac{1}{\sqrt{N}} \afrac{ \StDvYt }{\sqrt{ \VarXt }}
         \\
         \afrac{\VarXt}{\StDvXt^2} \ll \afrac{\VarYt}{\StDvYt^2}
         \arrforw
         \text{\mbox{OLS-x:y}}
         \spazio
         \ddd{ p_{\tz{x}} } \longrightarrow \afrac{1}{\sqrt{N}} \afrac{ \StDvXt }{\sqrt{ \VarYt }}
  \end{cases}
\punto
\end{gather}

\section{Case studies}
\label{pa:cases}

\subsection{B. C. Reed's example}
\label{pa:casesReed}

Consider B. C. Reed's example I~\cite{bi:Reed}, which was analyzed
in~\cite{bi:PHBorcherdstAndCVShetht} by using approximate expressions for the standard errors.
These data derive from a real, physical
situation: calibrating the colors of globular star clusters as a function of
their spectral types. The abscissa represents the difference between the
ultraviolet and yellow light magnitudes of the clusters and the ordinate
represents the difference between the yellow and infrared magnitudes.

\begin{figure}[htb]
  \centering
  \includegraphics[width=0.40\textwidth]{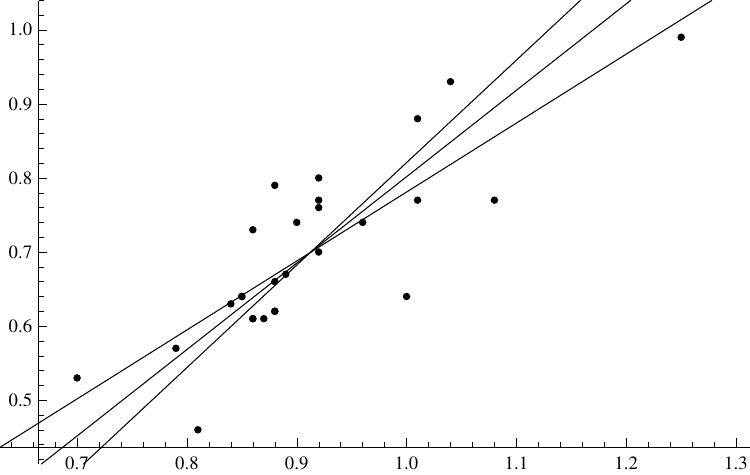}            
  \caption{Data points and the three best fit lines: OLS-y:x (the smallest slope), LSFSL-SWM (the intermediate slope) and OLS-x:y (the largest slope).}
\label{fi:BCReed1}
\end{figure}

The results are summarized in Figure~\ref{fi:BCReed1} and Table~\ref{ta:BCReed} .
The results for the LSFSL-SWM are in perfect agreement with the paper of the
author while the approximate treatment
of paper~\cite{bi:PHBorcherdstAndCVShetht} leads to slightly reduced error
estimates. Note that the correlation coefficient is missing in the expressions for the
standard errors, formula (10) of second paper in~\cite{bi:PHBorcherdstAndCVShetht},
but the authors do not use it; it is zero in the SWM.
On the other hand
equation (21) in the second paper in~\cite{bi:Reed} is correct in the commonly
encountered case that the errors in the two variables are uncorrelated;
unfortunately this is not explicitly stated in the text but only in end-note (9).

\begin{table}[htb]
\begin{tabular}{|c|c|c|}
\hline
OLS-y:x         & $ p_y = +0.9311 \pm 0.1299 $   & $ q_y = -0.1501 \pm 0.1192 $      \\
OLS-x:y         & $ p_y = +1.3839 \pm 0.1930 $   & $ q_y = -0.5631 \pm 0.1768 $      \\
LSFSL-SWM       & $ p_y = +1.1668 \pm 0.1704 $   & $ q_y = -0.3652 \pm 0.1561 $      \\
\hline
\end{tabular}
\caption{Best fit results (slope, intercept): OLS-y:x, LSFSL-SWM and OLS-x:y.}
\label{ta:BCReed}
\end{table}

Since the data are given with only two significant figures, the number of digits
quoted is not justified. However, following B. C. Reed~\cite{bi:Reed},
the point is to provide figures against which others can compare the results of
their own algorithms as if the original data are regarded as exact.

\begin{figure}[htb]
  \centering
  \includegraphics[width=0.40\textwidth]{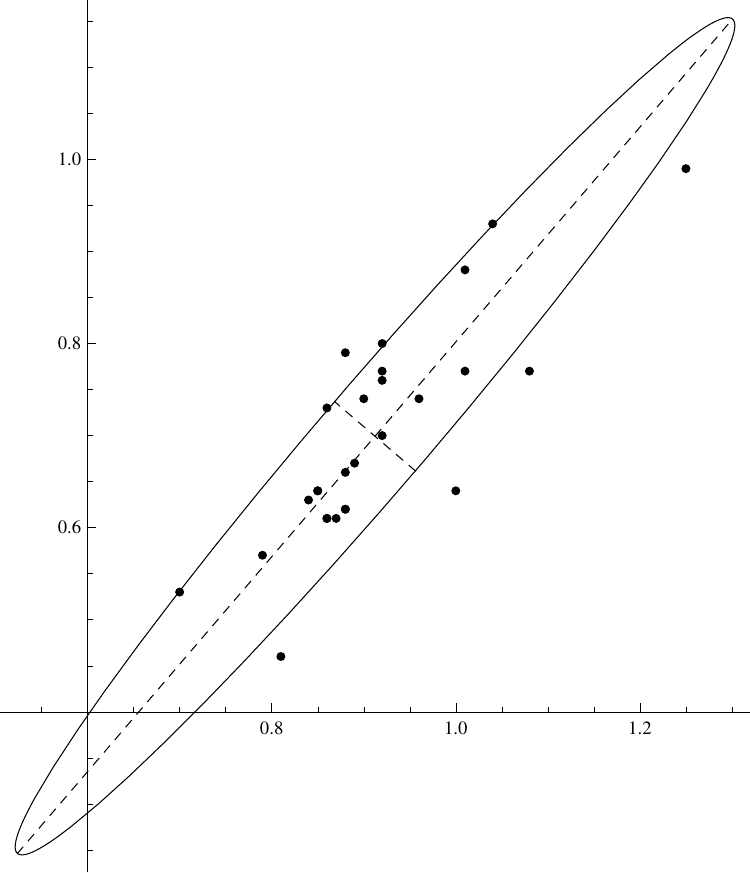}            
  \caption{Data points and results of PCA: the ellipse has its semi-axes along
    the directions of the eigenvectors and the semi-axes have the same length as the eigenvalues.}
\label{fi:BCReed2}
\end{figure}

The results of PCA are shown in Figure~\ref{fi:BCReed2}.

\subsection{A light-detector}

Consider a CCD-like light detector, where the image produced by a suitable optics is
focused (a camera, for instance). 
Suppose it must be used to identify straight light tracks, produced by a moving
point-like light source in the night sky,
for instance to observe meteors, airplanes or
extensive air-showers produced by cosmic radiation~\cite{bi:SEUSO}.
Let the typical length of the track be $\sigma_L$, defined as the standard
deviation of the data points along the track.

Assume, firstly, that the combination of the intrinsic spread of the
point-like light source and
the point-spread function of the optics gives, on the photo-detector plane, a standard deviation
$ \sigma_x = \sigma_y \equiv \scaleFact $ much larger than the pixel size, so that
binning effects can be neglected.
The expected angular resolution can be estimated as (equation~\ref{eq:ErrorThetaGoodCorrelation}):
\begin{equation}
       \ddd{\theta} \simeq
        \afrac{\scaleFact}{\sqrt{N}} \afrac{1}{ \sqrt{\VarCsi} } =
        \afrac{\scaleFact}{\sqrt{N}} \afrac{1}{ \sigma_L } 
\punto
\end{equation}

As a second example consider the case that the pixel size, $d$, is much larger
than the transverse size of the image track on 
the photo-detector plane.
The uncertainty on the measurements can be taken as $ \scaleFact = d/\sqrt{12} $,
assuming a uniform probability distribution inside the pixel. Assuming
uncorrelated measurements, the expected angular resolution can be estimated as (equation~\ref{eq:ErrorThetaGoodCorrelation}):
\begin{equation}
        \ddd{\theta} \simeq
        \afrac{\scaleFact}{\sqrt{N}} \afrac{1}{ \sqrt{\VarCsi} } =
        \afrac{1}{\sqrt{N}} \afrac{d}{\sqrt{12}} \afrac{1}{ \sigma_L } 
\punto
\end{equation}

Both analytical relations can be used to optimize the photo-detector parameters, given the 
desired angular resolution.

\subsection{Kinematics of a image track in a photo-detector}

The explicit analytical formula~\ref{eq:dtheta} was used, in
reference~\cite{bi:SEUSO}, in the design of a 
photo-detector to observe the image track of the extensive air-showers produced
in the atmosphere by ultra high energy cosmic radiation.  
The explicit analytical formulas, written in terms
of the photo-detector parameters, were then used to optimize the photo-detector
design.

\nn{

\section{Toy Monte Carlo simulations}
\label{pa:simul}

\subsection{Setup of the toy Monte Carlo simulations}

In order to cross-check and evaluate the accuracy of the formulas for the standard errors
derived in this paper, extensive Monte Carlo
simulations~\cite{bi:Wolfram} were carried on as follows. 

\begin{enumerate}

\item
Straight lines were simulated in the $xy$ plane, with uniformly distributed
random angle $\theta$ and a Gaussian distribution of the signed-distance, $c$, with zero mean
and a standard deviation of $D=1$.
For each straight line a fixed number, $N$, of true data points was simulated, on the
straight line, with a uniform random distribution along a 
segment of length $L=1$ of the straight line.

\item 
Each true data point was converted into a simulated measured data point
(i.e. simulated measurements)
according to a bi-dimensional Gaussian distribution centered at the true data
point and having equal fixed standard errors, 
$ \scaleFact \equiv \sigma_x = \sigma_y $, 
and zero correlation coefficient.  The best-fit straight line was then
calculated.

\item
For each simulated straight line and its set of true data points, the procedure at step 2) was repeated
for a number of times; each repetition is called \textit{one iteration}
and, typically, the number of iterations is $ n_I \approx 1000 $.  Statistics was then
collected of the fitted angle/signed-distance results, for one specific true
straight line and its set of true data points.

\item
The procedure at steps 1), 2) and 3) was repeated by simulating a number of different
straight lines; each simulation of one straight line is called \textit{one run} and, typically,
the number of runs is $ n_R \approx 1000 $. For all the runs,
statistics was collected of the fitted parameters with respect to the known true
parameters of the simulated straight line and its set of true data points.

\item
The procedure at steps 1), 2), 3) and 4) was then repeated for different
values of the number of measured data points, $N$, and of the standard error of
the single data point, $\scaleFact$. The different values for $N$
and $\scaleFact$ where chosen in such a way to keep an approximately constant
ratio between one value and the next one, inside a pre-defined fixed range.
Therefore different sets of parameters, $\pton{N,\scaleFact}$, were simulated, with:
$ 3 \leq N \leq 100 $ and $ 0.001 \leq \scaleFact \leq 1 $, as follows:

\begin{gather}
N=\pgra{ 3; 4; 5; 6; 8; 10; 15; 20; 30; 50; 100};
\\
\scaleFact :
\begin{cases}
\spazio\text{coarse-scan in $\scaleFact$:} \spazio \scaleFact = \pgra{ 1; 0.1; 0.01; 0.001 } \propto 10^{-k} 
\\
\spazio\text{fine-scan in $\scaleFact$:}   \spazio \scaleFact = \pgra{ 0.1; 0.0631; 0.0398; 0.0251; 0.0158; 0.01 } \propto 10^{-k/5}
\end{cases}
\label{eq:param}
\end{gather}

\end{enumerate}


The problem is invariant under a common rescaling of the two coordinates,
so that the results are expected to depend only on the ratio $\scaleFact/L$.

The largest simulated value of the data point error, $\scaleFact/L = 1$, would give so large errors with respect to the
length of the segment of the straight line that it is a non realistic case. It has been
simulated to verify that, in this case, the approximation of the standard formula for the propagation of errors
fails and to determine under what conditions the formula becomes an accurate estimate of the
standard error. 

It would be also possible to simulate straight lines with
absolute values of the signed distance, $ c \gtrsim L = 1 $.
However the standard error formulas, equations~\ref{eq:dtheta},~\ref{eq:dc}
and~\ref{eq:dctheta}, show that it is better to set the origin of the Cartesian
Coordinate System as close as possible to the centroid of the measured data
points, in order to have $c \simeq 0$, so that $Z \simeq 0$ and both the error on the signed
distance and the covariance between the angle and the signed-distance are minimized.
This is of course always possible, by means of a suitable translation, and therefore there is no need to simulate
larger absolute values of the signed-distance. 

For both the angle $\theta$ and the signed-distance $c$, the three following
quantities were calculated.

\begin{enumerate}

\item
The standard deviation of the distributions of the fitted angle and
signed-distance, $\est{\sigma}[\est{\theta}]$ and $\est{\sigma}[\est{c}]$, were 
calculated for every run; 
for any set of parameters the average was taken over all the runs:
$\av{\est{\sigma}[\est{\theta}]}$ and $\av{\est{\sigma}[\est{c}]}$.

This is what is usually defined as the standard deviation of the estimator.

\item
Equations~\ref{eq:dtheta} and~\ref{eq:dc} were used, for all iterations of every
run, to calculate the standard errors from the simulated measured data points
(i.e. simulated measurements) and
the median was taken over all iterations of every run, $M[\est{\ddd{\theta}}]$ and $M[\est{\ddd{c}}]$; 
for any set of parameters the average was taken over all the runs:
$\av{M[\est{\ddd{\theta}}]}$ and $\av{M[\est{\ddd{c}}]}$.

This is an estimate of the typical error one would calculate from a real set of data
points.
The median is used in order to provide a robust estimation, as outlying large
values may show-up for certain combinations of true data-points (for instance
all true data-points clustered one close to the other).

\item
For each run the standard errors were 
computed from the true data points via equations~\ref{eq:dtheta},
for $\ddd{\theta}_0$, and~\ref{eq:dc}, for $\ddd{c}_0$; 
for any set of parameters the average was taken over all the runs:
$\av{\ddd{\theta}_0}$ and $\av{\ddd{c}_0}$. 

For any given straight line and any set of true data-points on it,
$\ddd{\theta}_0$ and $\ddd{c}_0$ are the expected standard errors
calculated by the standard formula for the propagation of errors.
Therefore $\av{\ddd{\theta}_0}$ and $\av{\ddd{c}_0}$ are a sort of reference values, setting the expected magnitude of the
standard errors, for any given set of parameters, and useful as normalization
factors to compare different set of parameters.

\end{enumerate}

Some of the results obtained by the extensive Monte Carlo simulations are summarized in the rest of this section.

\subsection{Results of the toy Monte Carlo simulations}

The results of the different simulations were re-normalized by $\scaleFact$, in order
to compare results for different values of $\scaleFact$,
and are shown as a function of $N$ for different values of $\scaleFact/L$.

\subsubsection{Results for \texorpdfstring{$\av{\ddd{\theta}_0}$}{.....} and \texorpdfstring{$\av{\ddd{c}_0}$}{.....}}

All the simulations for $\av{\ddd{\theta}_0}$ and $\av{\ddd{c}_0}$, 
after normalizing for $\scaleFact$,
show a very similar behavior as a function of $N$, 
for the different values of $\scaleFact/L$, 
as shown in Figure~\ref{fig:ResMCZero}.

The behavior as a function of $N$ is fitted 
by $\sim 1/\sqrt{N+4}$, for $\av{\ddd{\theta}_0}$, and 
by $\sim 1/\sqrt{N+1}$, for $\av{\ddd{c}_0}$
(the result of the fit not shown in the Figures).

\begin{figure}[htb]
\begin{center}
  \centering
  \subfloat[   Angle $\theta$.  ]{\includegraphics[width=0.48\textwidth]{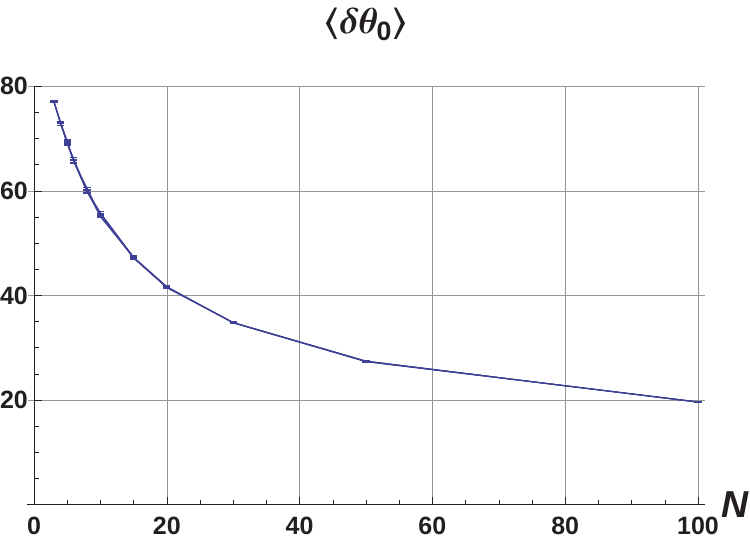}}
\spazio
  \centering
  \subfloat[Signed-distance $c$.]{\includegraphics[width=0.48\textwidth]{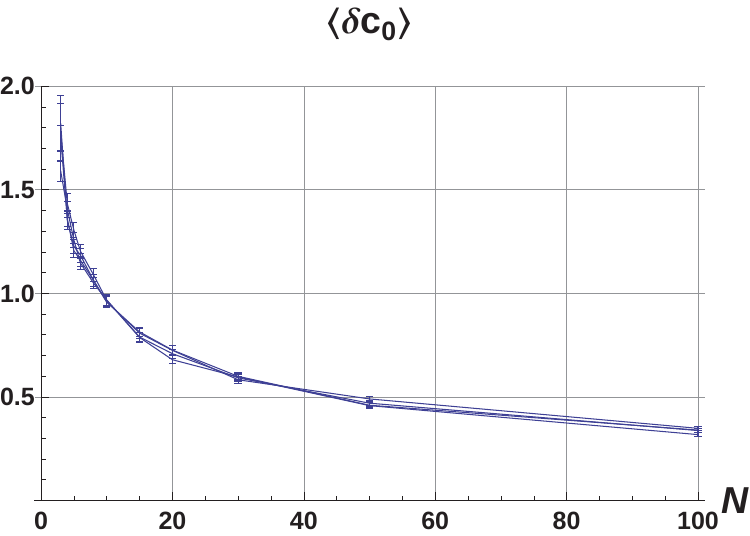}}
\end{center}
\caption{Results of the simulations for $\av{\ddd{\theta}_0}$ and $\av{\ddd{c}_0}$,
         as a function of $N$, 
         for $\scaleFact/L=\pgra{1;0.1;0.01;0.001}$.
         The four curves on each plot are well superimposed and barely distinguishable. 
         Errors bars are very small and barely visible as well.} 
\label{fig:ResMCZero}
\end{figure}

\clearpage

\subsubsection{Results for \texorpdfstring{$\av{\est{\sigma}[\est{\theta}]}$}{.....} and \texorpdfstring{$\av{\est{\sigma}[\est{c}]}$}{.....}}

All the simulations for $\av{\est{\sigma}[\est{\theta}]}$ and $\av{\est{\sigma}[\est{c}]}$,
after normalizing for $\scaleFact$,
show a very similar behavior as a function of $N$, 
for the different values of $\scaleFact/L \lesssim 0.05$,
as shown individually in Figure~\ref{fig:ResMCSig1} and all together in Figure~\ref{fig:ResMCSig1bis}.

Moreover, for values $\scaleFact/L \lesssim 0.05$,
the behavior as a function of $N$ is fitted 
by $ \sim 1/\sqrt{N+4} $, for $\av{\est{\sigma}[\est{\theta}]}$, 
and by $ \sim 1/\sqrt{N+1} $, for $\av{\est{\sigma}[\est{c}]}$ 
(the result of the fit not shown in the Figures).

\begin{figure}[htp] 
\begin{center}
  \centering
  \subfloat[   Angle $\theta$.  ]{\includegraphics[width=0.35\textwidth]{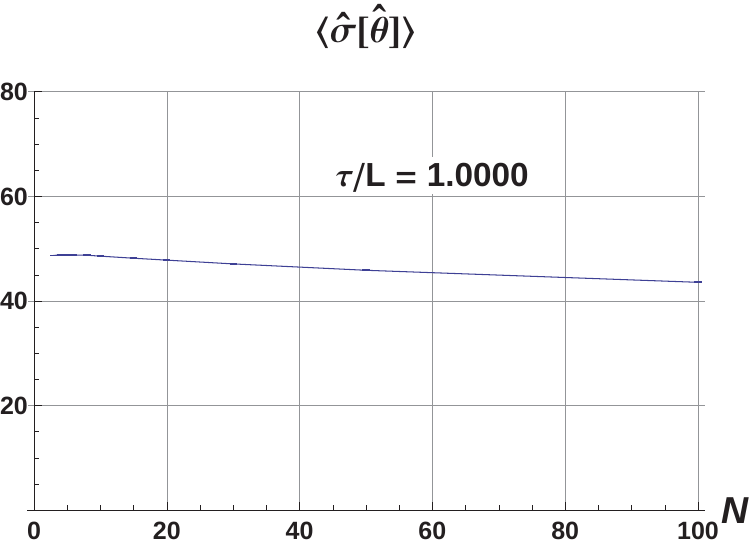}}
\spazio
  \centering
  \subfloat[Signed-distance $c$.]{\includegraphics[width=0.35\textwidth]{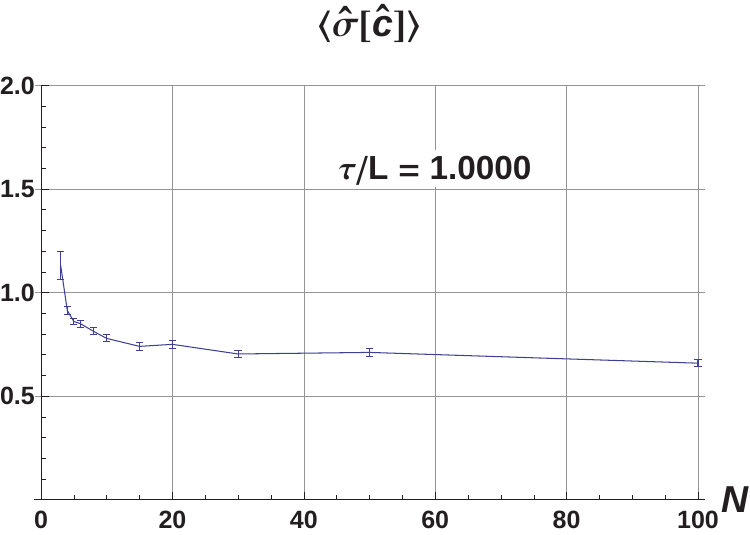}}
\\
  \centering
  \subfloat[   Angle $\theta$.  ]{\includegraphics[width=0.35\textwidth]{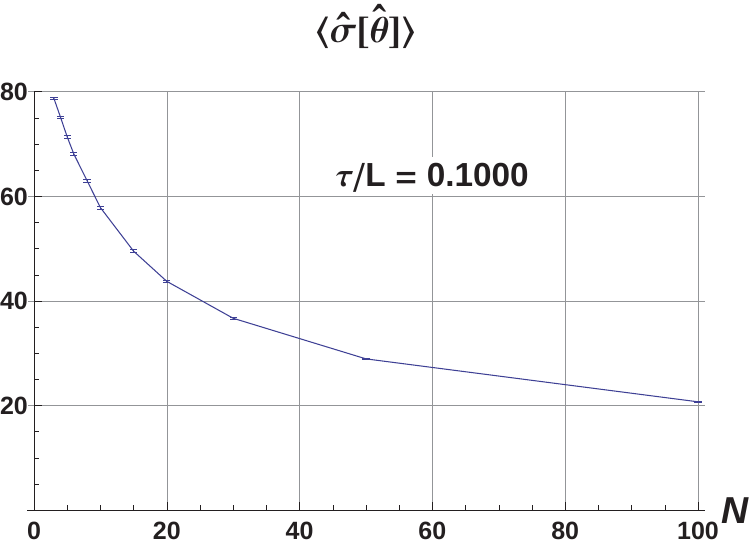}}
\spazio
  \centering
  \subfloat[Signed-distance $c$.]{\includegraphics[width=0.35\textwidth]{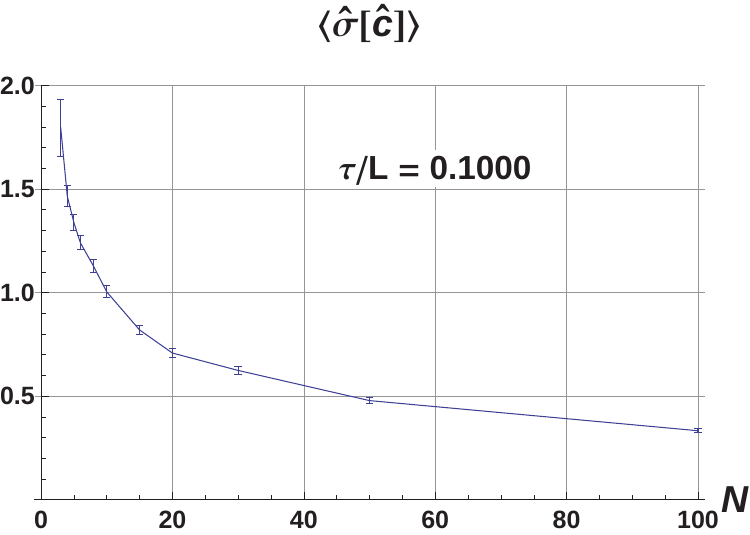}}
\\
  \centering
  \subfloat[   Angle $\theta$.  ]{\includegraphics[width=0.35\textwidth]{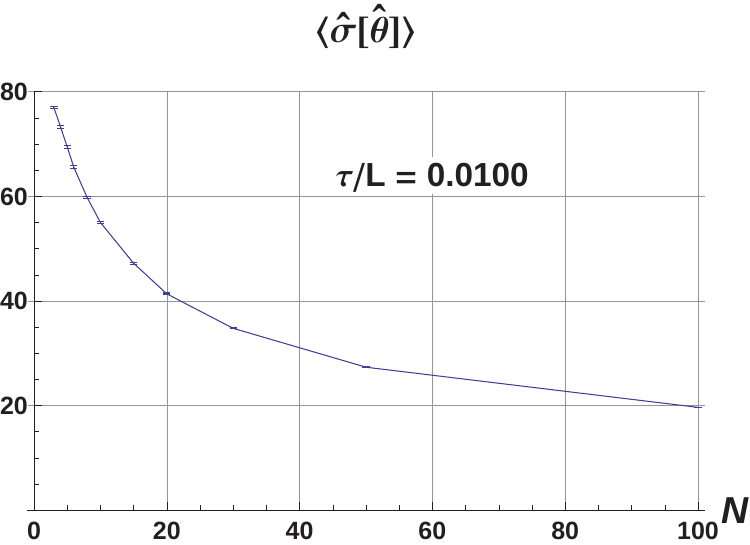}}
\spazio
  \centering
  \subfloat[Signed-distance $c$.]{\includegraphics[width=0.35\textwidth]{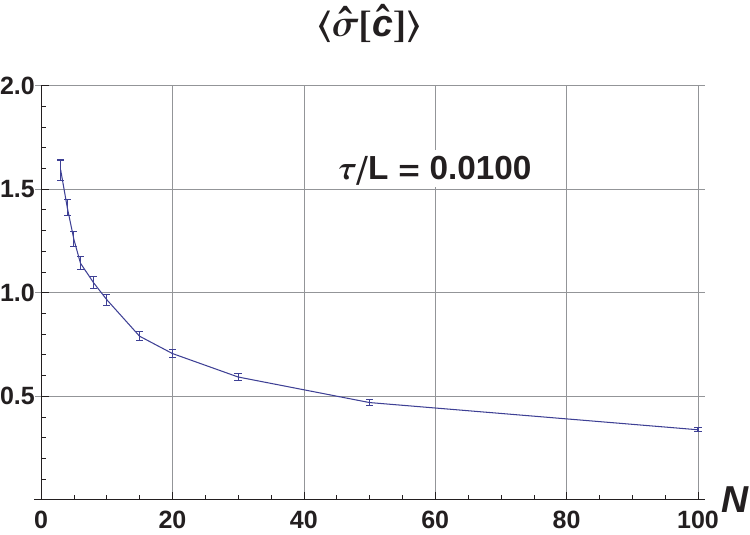}}
\\
  \centering
  \subfloat[   Angle $\theta$.  ]{\includegraphics[width=0.35\textwidth]{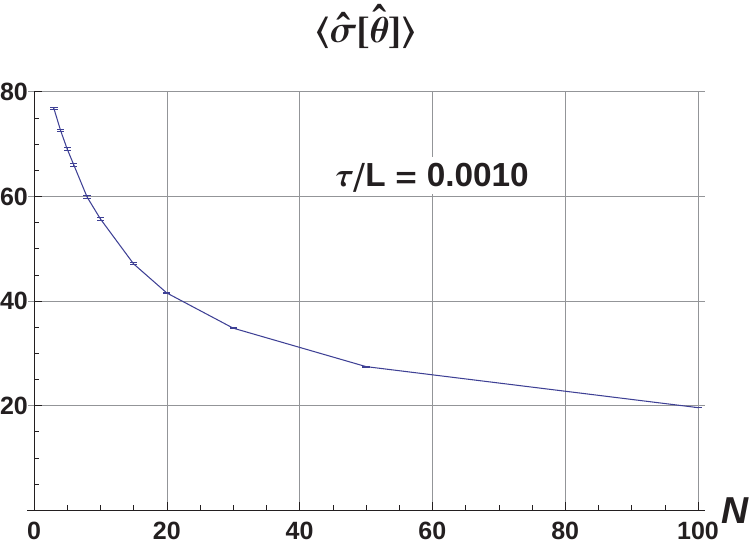}}
\spazio
  \centering
  \subfloat[Signed-distance $c$.]{\includegraphics[width=0.35\textwidth]{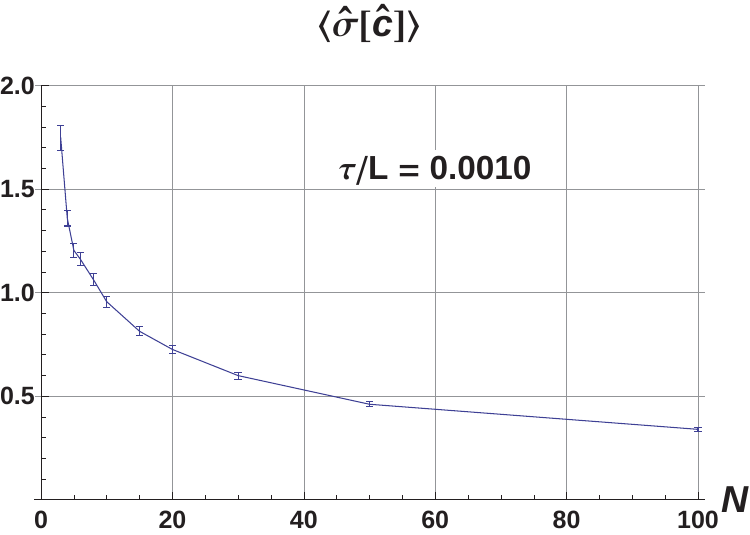}}
\end{center}  
\caption{Results of the simulations for $\av{\est{\sigma}[\est{\theta}]}$ and $\av{\est{\sigma}[\est{c}]}$,
         as a function of $N$,
         for $\scaleFact/L=\pgra{1;0.1;0.01;0.001}$.
         Errors bars are very small and barely visible.}
\label{fig:ResMCSig1}
\end{figure}

\clearpage

\begin{figure}[htb] 
\begin{center}
  \centering
  \subfloat[   Angle $\theta$.  ]{\includegraphics[width=0.48\textwidth]{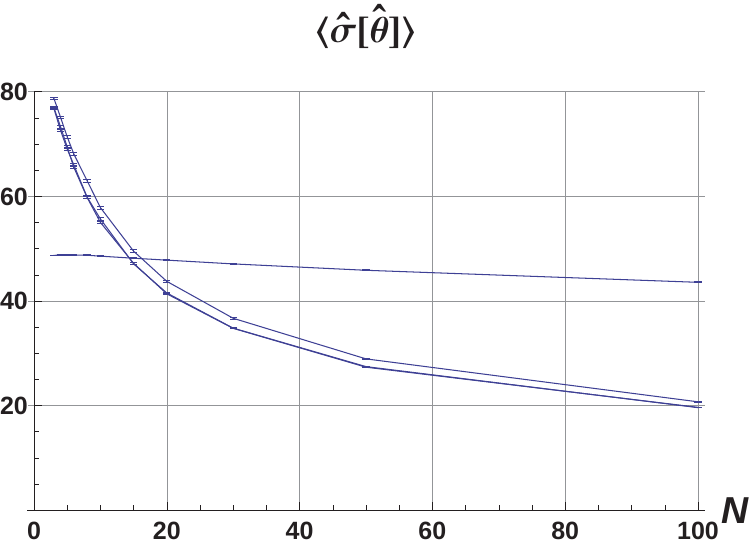}}
\spazio
  \centering
  \subfloat[Signed-distance $c$.]{\includegraphics[width=0.48\textwidth]{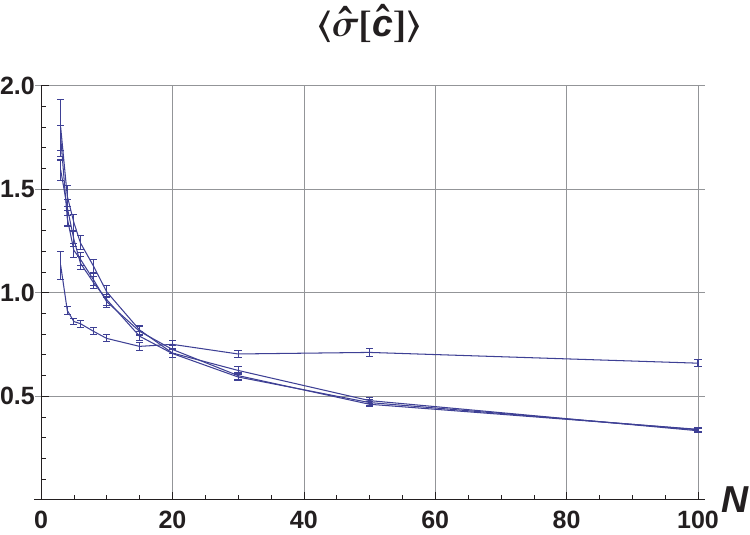}}
\end{center}  
\caption{Results of the simulations for $\av{\est{\sigma}[\est{\theta}]}$ and $\av{\est{\sigma}[\est{c}]}$,
         as a function of $N$,
         for $\scaleFact/L=\pgra{1;0.1;0.01;0.001}$.
         The two almost horizontal curves on each plot correspond to $\scaleFact/L=1$.
         The two curves with the smallest $\scaleFact/L$ are well superimposed and barely distinguishable. 
         Errors bars are very small and barely visible.}
\label{fig:ResMCSig1bis}
\end{figure}

In order to study the behavior for $ 0.01 \lesssim \scaleFact/L \lesssim 0.1 $,
a finer scan has been done in this interval;
the plots are individually shown 
in Figure~\ref{fig:ResMCSig2}, for $ \scaleFact/L = \pgra{  0.0631; 0.0398; 0.0251; 0.0158 }$,
and all together 
in Figure~\ref{fig:ResMCSig2bis}, for $ \scaleFact/L = \pgra{ 0.1; 0.0631; 0.0398; 0.0251; 0.0158; 0.01 }$.

\begin{figure}[htp] 
\begin{center}
  \centering
  \subfloat[   Angle $\theta$.  ]{\includegraphics[width=0.35\textwidth]{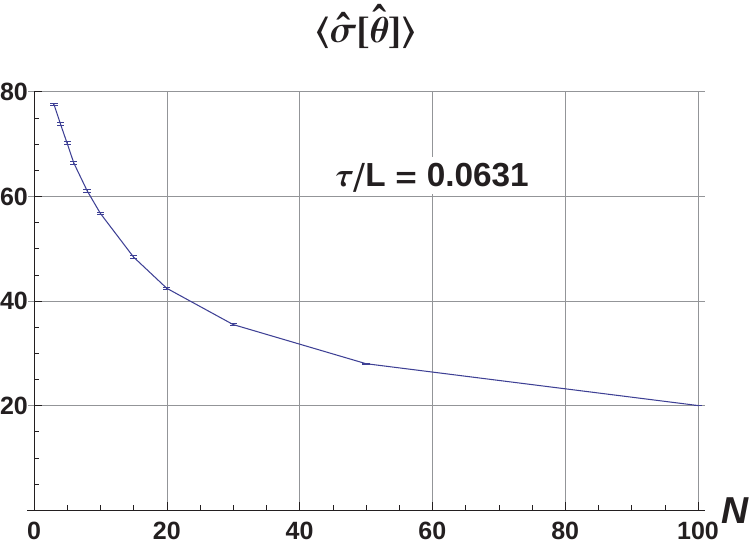}}
\spazio
  \centering
  \subfloat[Signed-distance $c$.]{\includegraphics[width=0.35\textwidth]{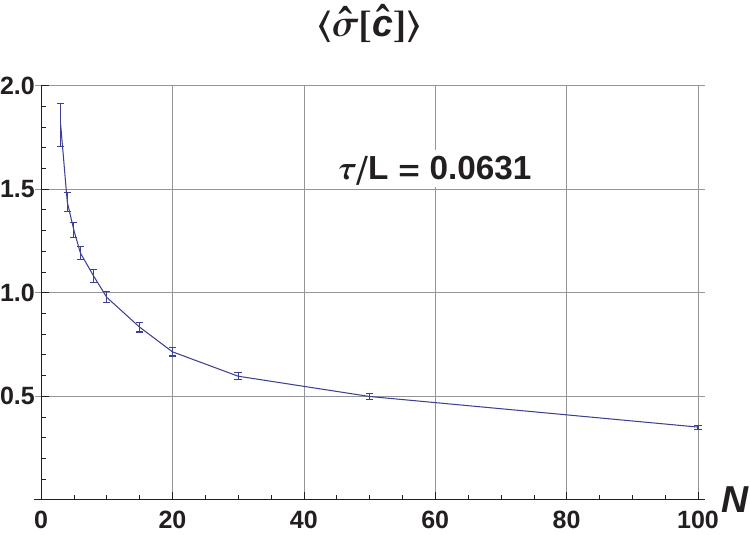}}
\\
  \centering
  \subfloat[   Angle $\theta$.  ]{\includegraphics[width=0.35\textwidth]{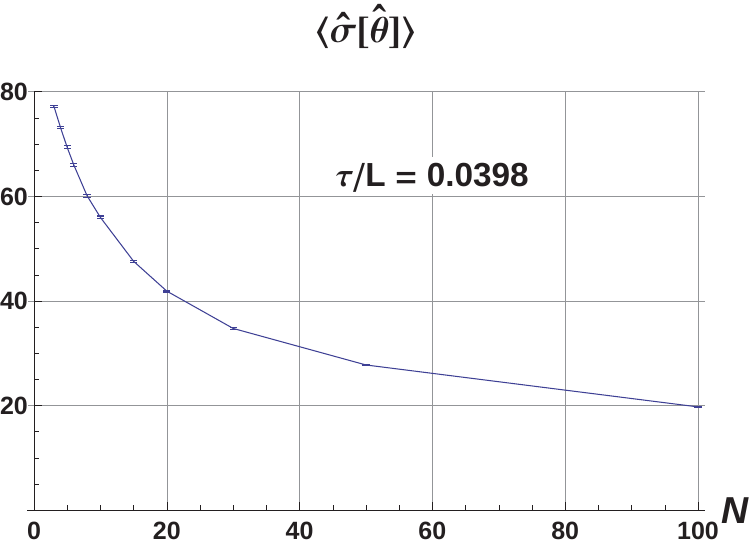}}
\spazio
  \centering
  \subfloat[Signed-distance $c$.]{\includegraphics[width=0.35\textwidth]{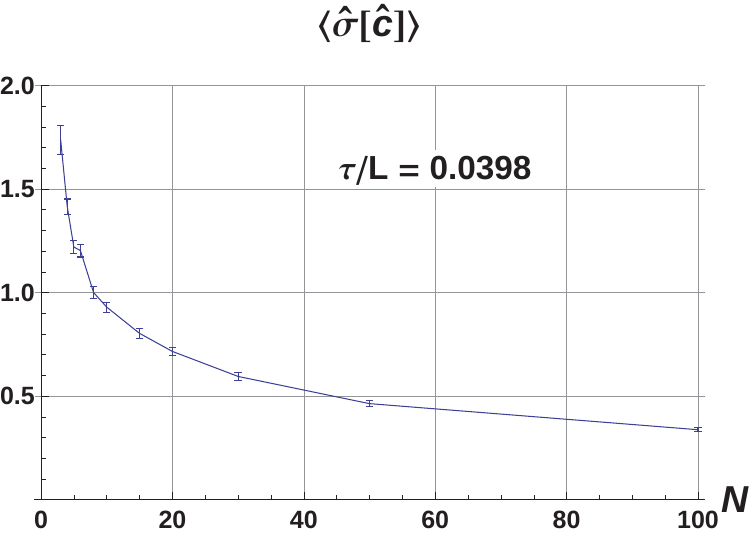}}
\\
  \centering
  \subfloat[   Angle $\theta$.  ]{\includegraphics[width=0.35\textwidth]{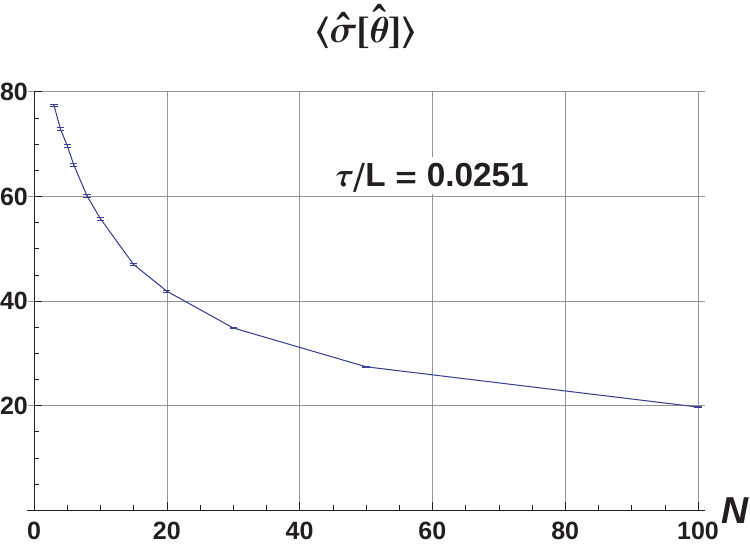}}
\spazio
  \centering
  \subfloat[Signed-distance $c$.]{\includegraphics[width=0.35\textwidth]{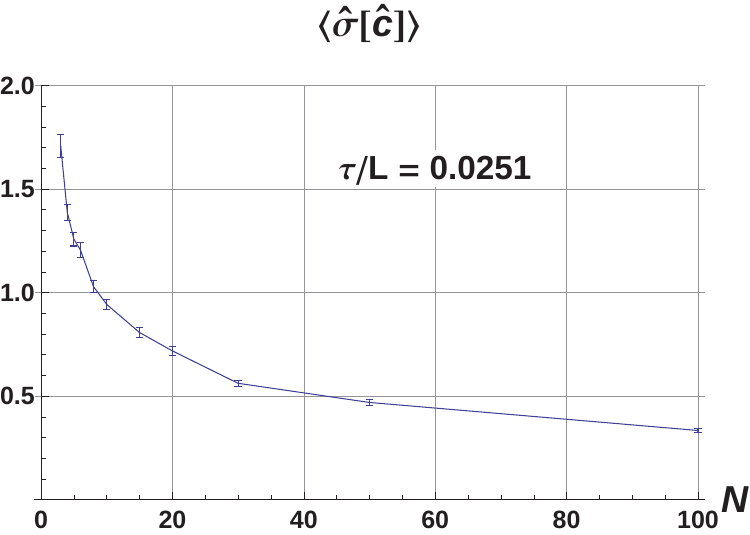}}
\\
  \centering
  \subfloat[   Angle $\theta$.  ]{\includegraphics[width=0.35\textwidth]{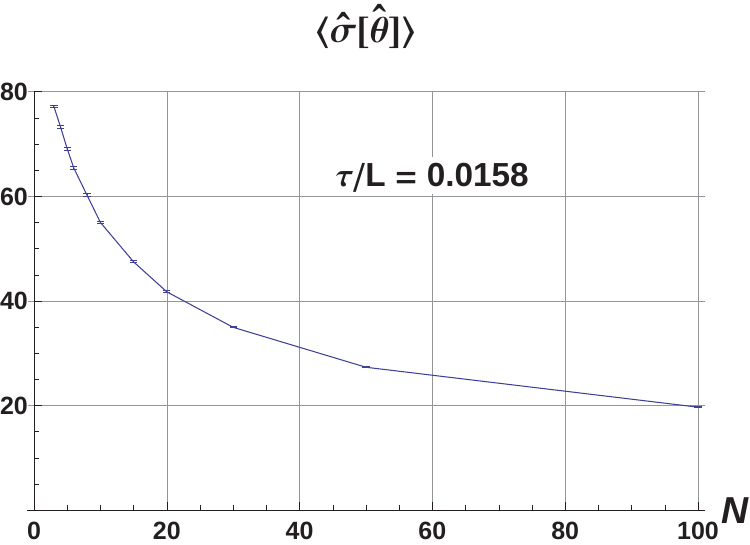}}
\spazio
  \centering
  \subfloat[Signed-distance $c$.]{\includegraphics[width=0.35\textwidth]{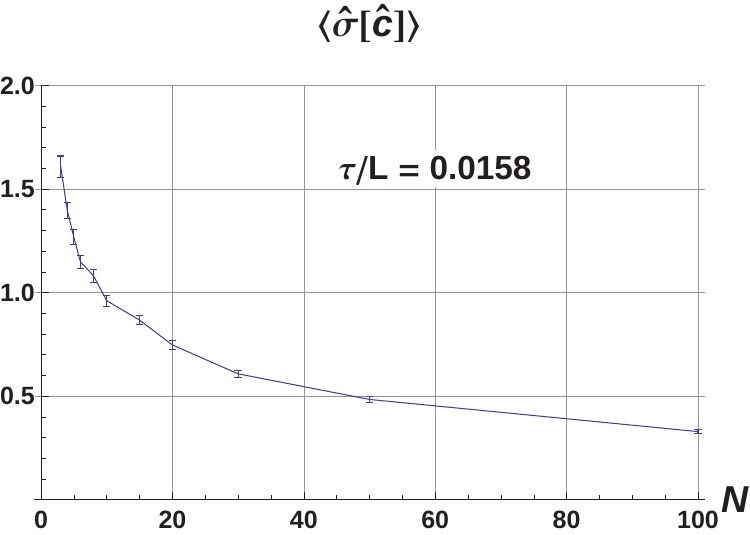}}
\end{center}  
\caption{Results of the simulations for $\av{\est{\sigma}[\est{\theta}]}$ and $\av{\est{\sigma}[\est{c}]}$,
         as a function of $N$,
         for $ \scaleFact/L = \pgra{ 0.0631; 0.0398; 0.0251; 0.0158 }$.
         Errors bars are very small and barely visible.}
\label{fig:ResMCSig2}
\end{figure}

\begin{figure}[htb] 
\begin{center}
  \centering
  \subfloat[   Angle $\theta$.  ]{\includegraphics[width=0.48\textwidth]{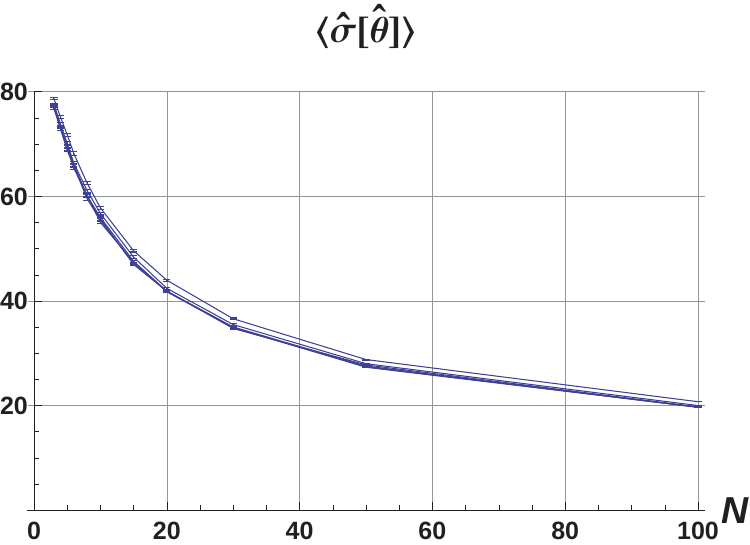}}
\spazio
  \centering
  \subfloat[Signed-distance $c$.]{\includegraphics[width=0.48\textwidth]{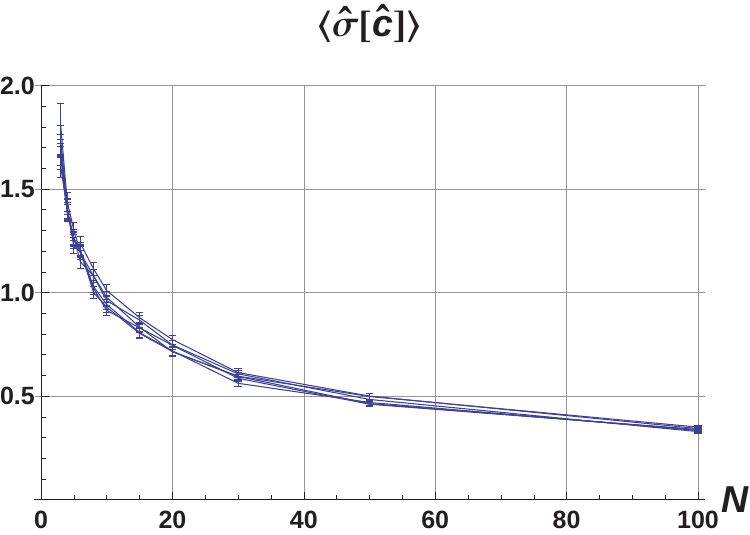}}
\end{center}  
\caption{Results of the simulations for $\av{\est{\sigma}[\est{\theta}]}$ and $\av{\est{\sigma}[\est{c}]}$,
         as a function of $N$,
         for $ \scaleFact/L = \pgra{ 0.1; 0.0631; 0.0398; 0.0251; 0.0158; 0.01 }$.
         Errors bars are very small and barely visible.}
\label{fig:ResMCSig2bis}
\end{figure}

Afterward, 
in order to get rid of the variability associated with the random straight line
and random set of true data-points on the straight line, the values 
of $\av{\est{\sigma}[\est{\theta}]}$ and $\av{\est{\sigma}[\est{c}]}$
were normalized to $\av{\ddd{\theta}_0}$ and $\av{\ddd{c}_0}$.

For $\scaleFact \lesssim 0.02$, no statistically significant
difference with $\av{\ddd{\theta}_0}$ and $\av{\ddd{c}_0}$ has been found and the ratios 
$\av{\est{\sigma}[\est{\theta}]}/\av{\ddd{\theta}_0}$ and $\av{\est{\sigma}[\est{c}]}/\av{\ddd{c}_0}$ 
are compatible with one, as shown in Figure~\ref{fig:ResMCSig3}.

\begin{figure}[htb] 
\begin{center}
  \centering
  \subfloat[   Angle $\theta$.  ]{\includegraphics[width=0.48\textwidth]{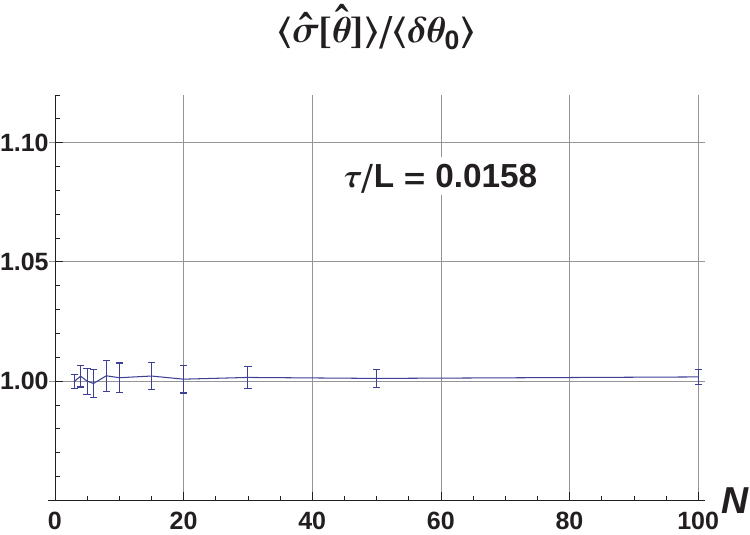}}
\spazio
  \centering
  \subfloat[Signed-distance $c$.]{\includegraphics[width=0.48\textwidth]{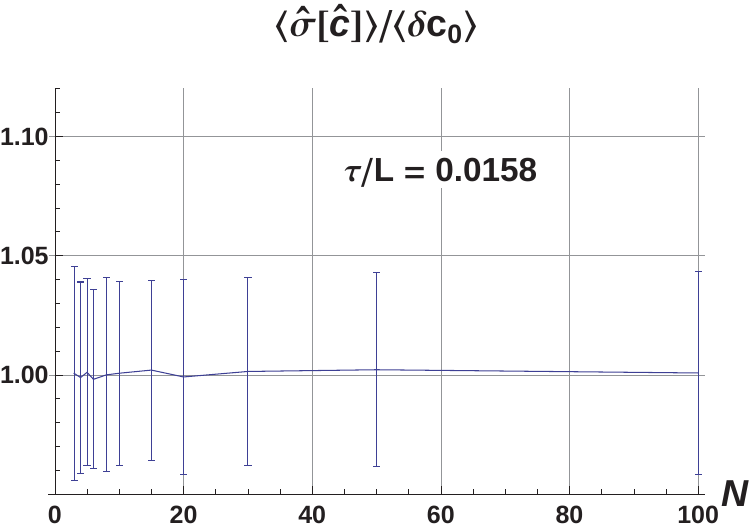}}
\end{center}  
\caption{Results of the simulations for the ratios $\av{\est{\sigma}[\est{\theta}]}/\av{\ddd{\theta}_0}$ and $\av{\est{\sigma}[\est{c}]}/\av{\ddd{c}_0}$,
         as a function of $N$,
         for $\scaleFact/L=0.0158$.}
\label{fig:ResMCSig3}
\end{figure}

On the other hand the behavior of the ratios 
$\av{\est{\sigma}[\est{\theta}]}/\av{\ddd{\theta}_0}$ and $\av{\est{\sigma}[\est{c}]}/\av{\ddd{c}_0}$
for values of $\scaleFact/L \gtrsim 0.02$,
as shown in Figure~\ref{fig:ResMCSig4}, 
shows a significant departure from the expected
$\av{\ddd{\theta}_0}$ and $\av{\ddd{c}_0}$, but not larger than $ \approx 5\%$ for $\scaleFact \lesssim 0.1$.

As a general trend, the ratios tend to one as $\scaleFact/L $ decreases, as
expected thanks to the improvement of the approximations made to derive the standard formula for the
propagation of errors.
Moreover the ratios tend to one at small $N$ anyway because at small $N$ the values of
the numerator and denominator became large with respect to their difference.

\begin{figure}[htb] 
\begin{center}
  \centering
  \subfloat[   Angle $\theta$.  ]{\includegraphics[width=0.48\textwidth]{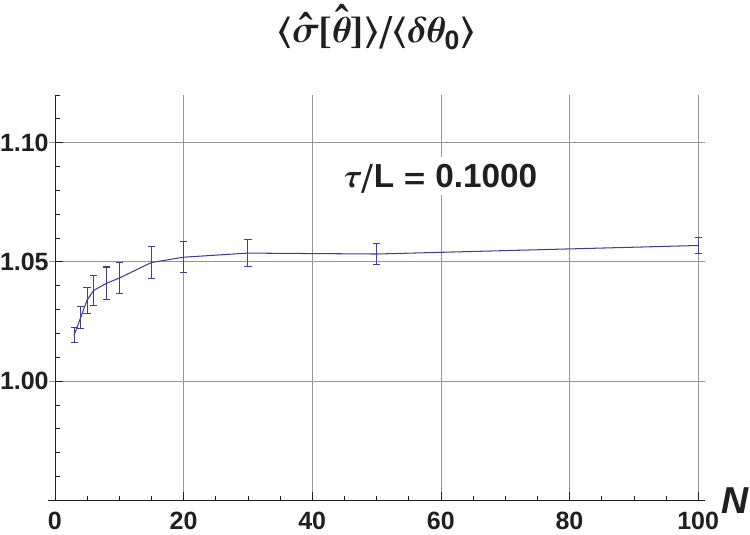}}
\spazio
  \centering
  \subfloat[Signed-distance $c$.]{\includegraphics[width=0.48\textwidth]{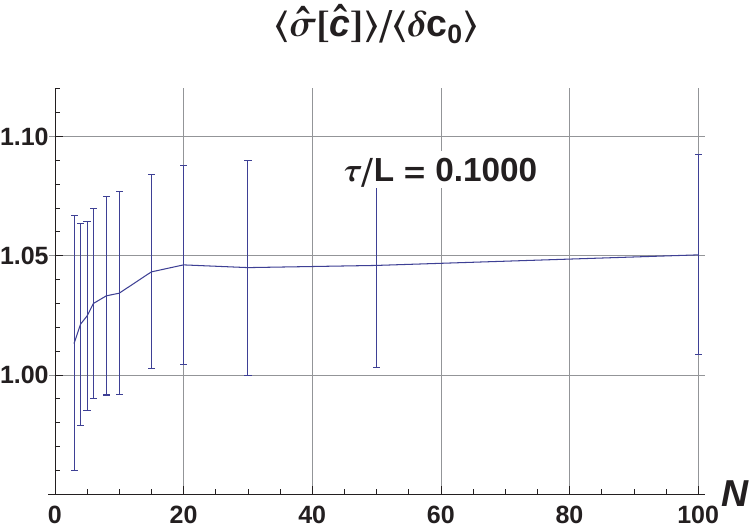}}
\end{center}  
\caption{Results of the simulations for the ratios $\av{\est{\sigma}[\est{\theta}]}/\av{\ddd{\theta}_0}$ and $\av{\est{\sigma}[\est{c}]}/\av{\ddd{c}_0}$,
         as a function of $N$,
         for $\scaleFact/L=0.1$.}
\label{fig:ResMCSig4}
\end{figure}

\clearpage
\subsubsection{Results for \texorpdfstring{$\av{M[\est{\ddd{\theta}}]}$}{.....} and \texorpdfstring{$\av{M[\est{\ddd{c}}]}$}{.....}}

All the simulations for $\av{M[\est{\ddd{\theta}}]}$ and $\av{M[\est{\ddd{c}}]}$,
after normalizing for $\scaleFact$,
show a very similar behavior as a function of $N$, 
for the different values of $\scaleFact/L \lesssim 0.05$,
as shown individually in Figure~\ref{fig:ResMCMed1} and all together in Figure~\ref{fig:ResMCMed1bis}.

Moreover, for values $\scaleFact/L \lesssim 0.05$,
the behavior as a function of $N$ is fitted
by $\sim 1/\sqrt{N+4}$, for $\av{M[\est{\ddd{\theta}}]}$, and 
by $\sim 1/\sqrt{N+1}$, for $\av{M[\est{\ddd{c}}]}$
(the result of the fit not shown in the Figures).

\begin{figure}[htp] 
\begin{center}
  \centering
  \subfloat[   Angle $\theta$.  ]{\includegraphics[width=0.35\textwidth]{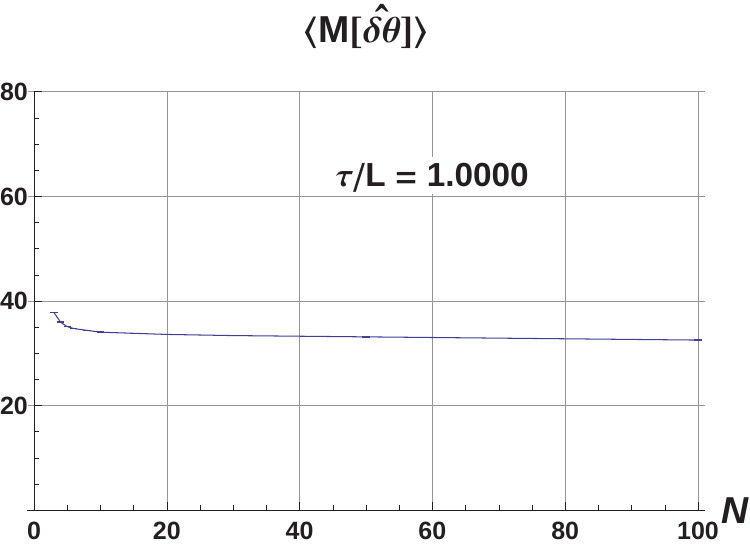}}
\spazio
  \centering
  \subfloat[Signed-distance $c$.]{\includegraphics[width=0.35\textwidth]{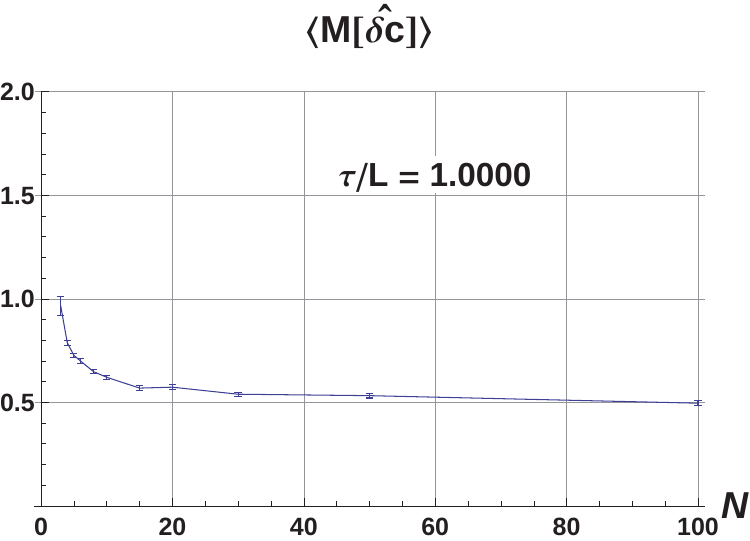}}
\\
  \centering
  \subfloat[   Angle $\theta$.  ]{\includegraphics[width=0.35\textwidth]{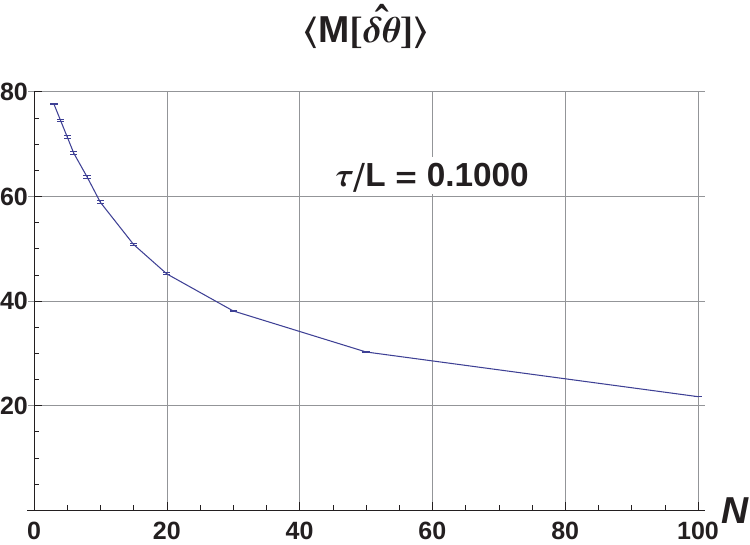}}
\spazio
  \centering
  \subfloat[Signed-distance $c$.]{\includegraphics[width=0.35\textwidth]{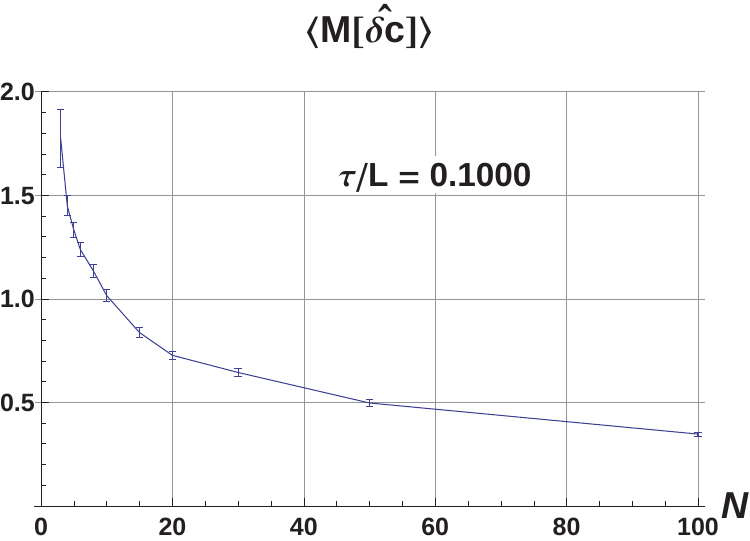}}
\\
  \centering
  \subfloat[   Angle $\theta$.  ]{\includegraphics[width=0.35\textwidth]{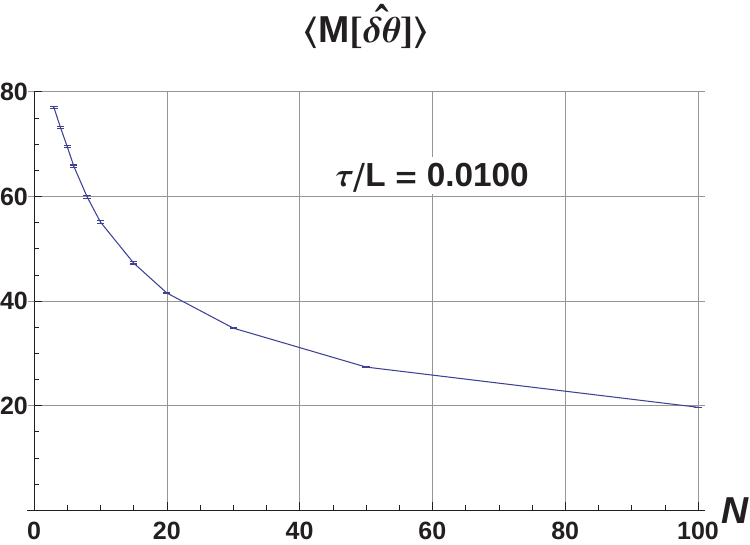}}
\spazio
  \centering
  \subfloat[Signed-distance $c$.]{\includegraphics[width=0.35\textwidth]{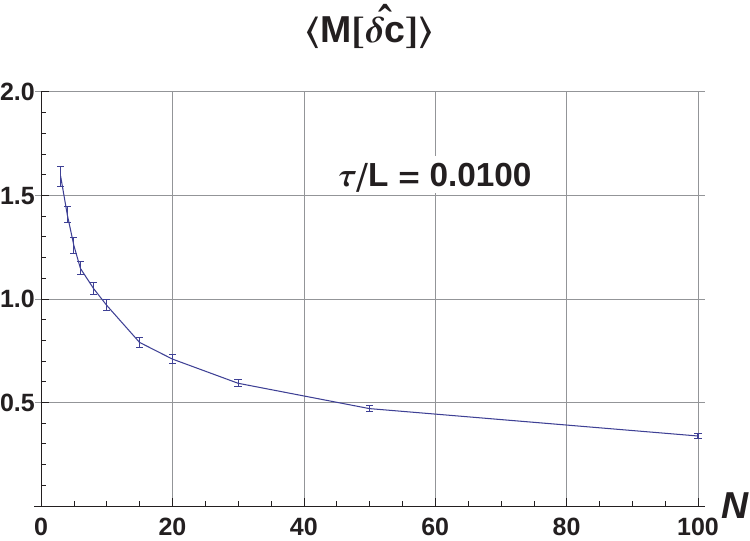}}
\\
  \centering
  \subfloat[   Angle $\theta$.  ]{\includegraphics[width=0.35\textwidth]{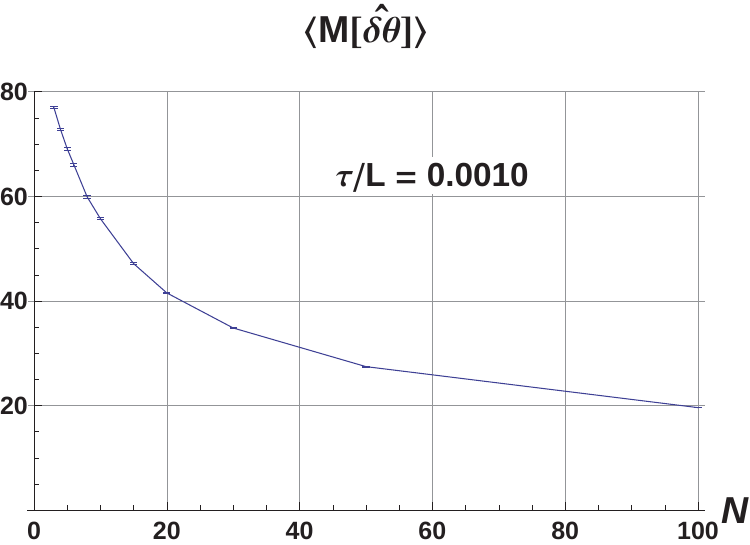}}
\spazio
  \centering
  \subfloat[Signed-distance $c$.]{\includegraphics[width=0.35\textwidth]{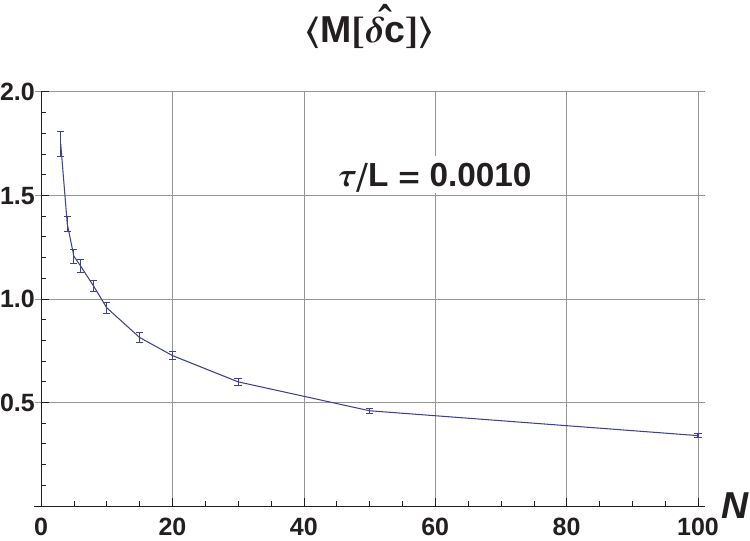}}
\end{center}  
\caption{Results of the simulations for $\av{M[\est{\ddd{\theta}}]}$ and $\av{M[\est{\ddd{c}}]}$,
         as a function of $N$,
         for $\scaleFact/L=\pgra{1;0.1;0.01;0.001}$.
         Errors bars are very small and barely visible.}
\label{fig:ResMCMed1}
\end{figure}

\begin{figure}[htb] 
\begin{center}
  \centering
  \subfloat[   Angle $\theta$.  ]{\includegraphics[width=0.48\textwidth]{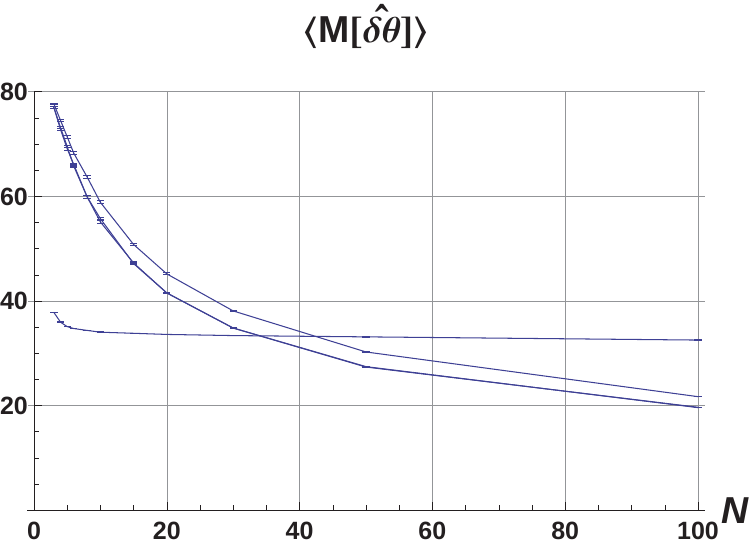}}
\spazio
  \centering
  \subfloat[Signed-distance $c$.]{\includegraphics[width=0.48\textwidth]{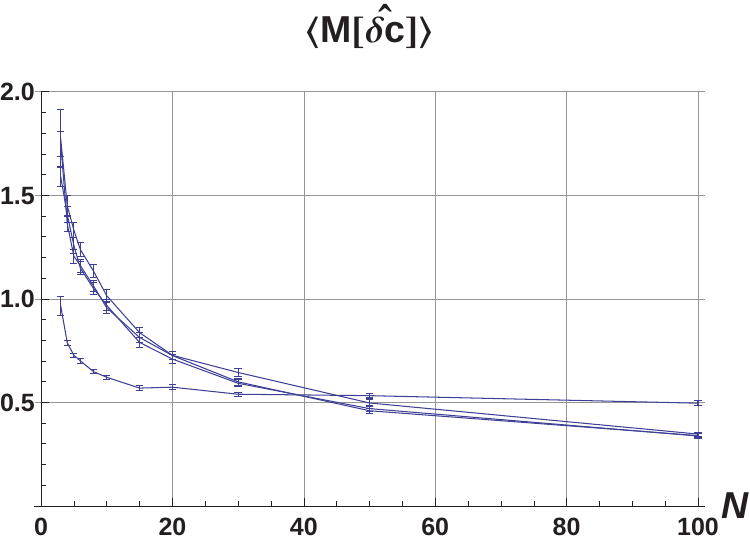}}
\end{center}  
\caption{Results of the simulations for $\av{M[\est{\ddd{\theta}}]}$ and $\av{M[\est{\ddd{c}}]}$,
         as a function of $N$,
         for $\scaleFact/L=\pgra{1;0.1;0.01;0.001}$.
         The two almost horizontal curves on each plot correspond to $\scaleFact/L=1$.
         The two curves with the smallest $\scaleFact/L$ are well superimposed and barely distinguishable. 
         Errors bars are very small and barely visible.}
\label{fig:ResMCMed1bis}
\end{figure}

In order to study the behavior for $ 0.01 \lesssim \scaleFact/L \lesssim 0.1 $,
a finer scan has been done in this interval;
the plots are individually shown
in Figure~\ref{fig:ResMCMed2}, for $ \scaleFact/L = \pgra{  0.0631; 0.0398; 0.0251; 0.0158 }$,
and all together
in Figure~\ref{fig:ResMCMed2bis}, for $ \scaleFact/L = \pgra{ 0.1; 0.0631; 0.0398; 0.0251; 0.0158; 0.01 }$.

\begin{figure}[htp] 
\begin{center}
  \centering
  \subfloat[   Angle $\theta$.  ]{\includegraphics[width=0.35\textwidth]{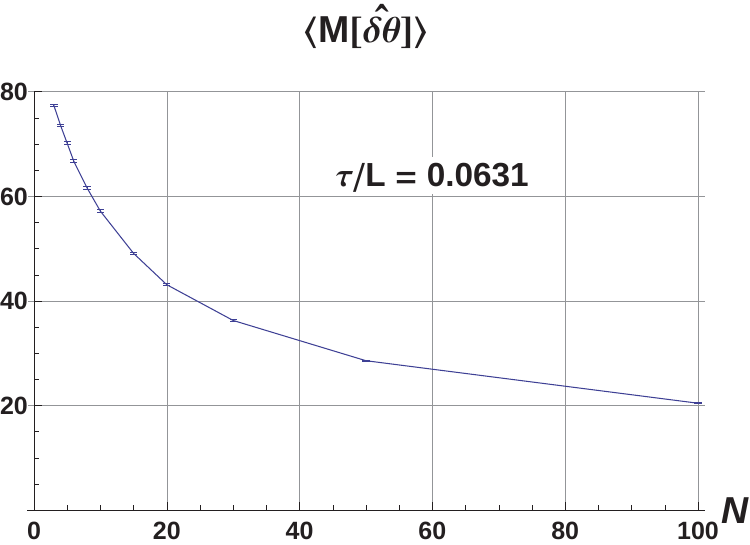}}
\spazio
  \centering
  \subfloat[Signed-distance $c$.]{\includegraphics[width=0.35\textwidth]{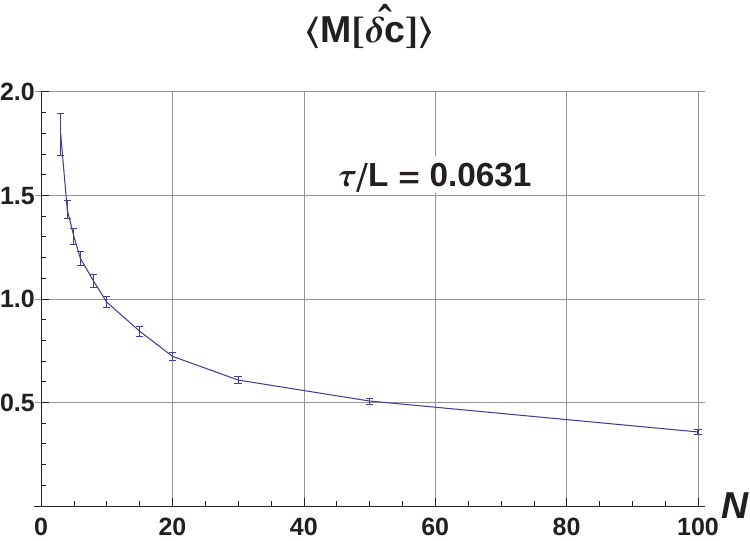}}
\\
  \centering
  \subfloat[   Angle $\theta$.  ]{\includegraphics[width=0.35\textwidth]{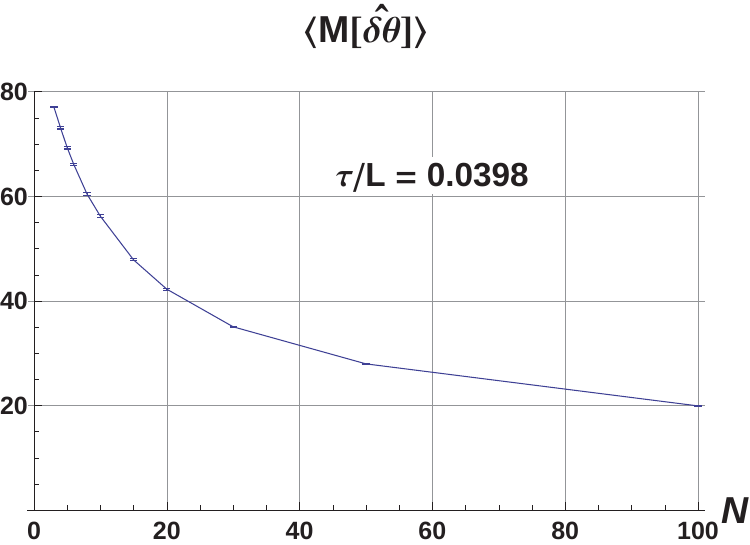}}
\spazio
  \centering
  \subfloat[Signed-distance $c$.]{\includegraphics[width=0.35\textwidth]{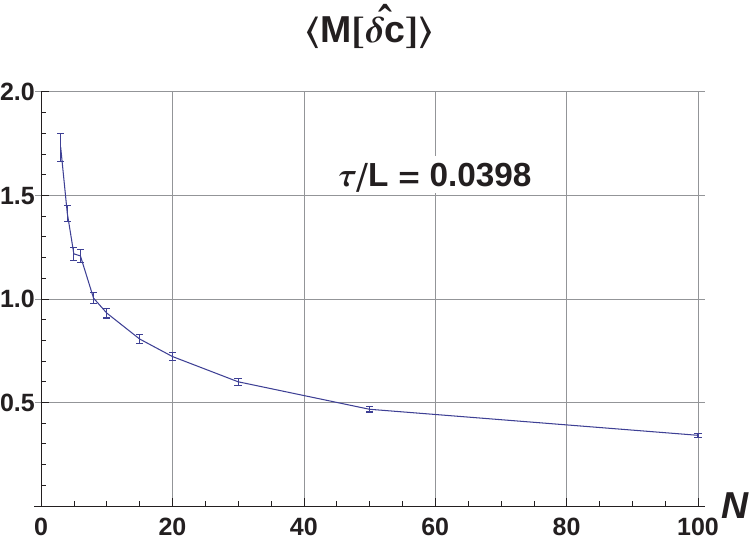}}
\\
  \centering
  \subfloat[   Angle $\theta$.  ]{\includegraphics[width=0.35\textwidth]{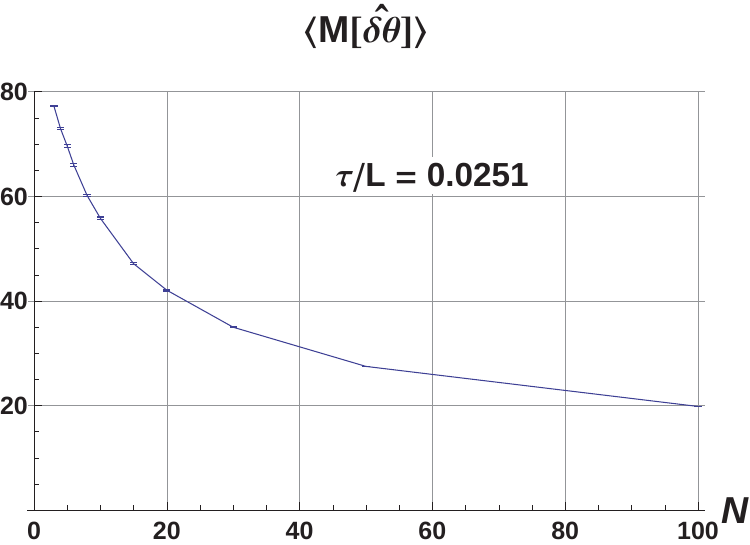}}
\spazio
  \centering
  \subfloat[Signed-distance $c$.]{\includegraphics[width=0.35\textwidth]{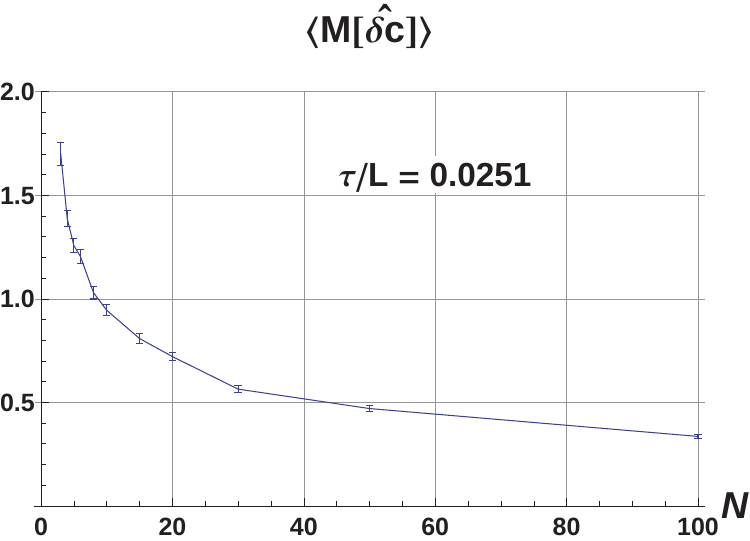}}
\\
  \centering
  \subfloat[   Angle $\theta$.  ]{\includegraphics[width=0.35\textwidth]{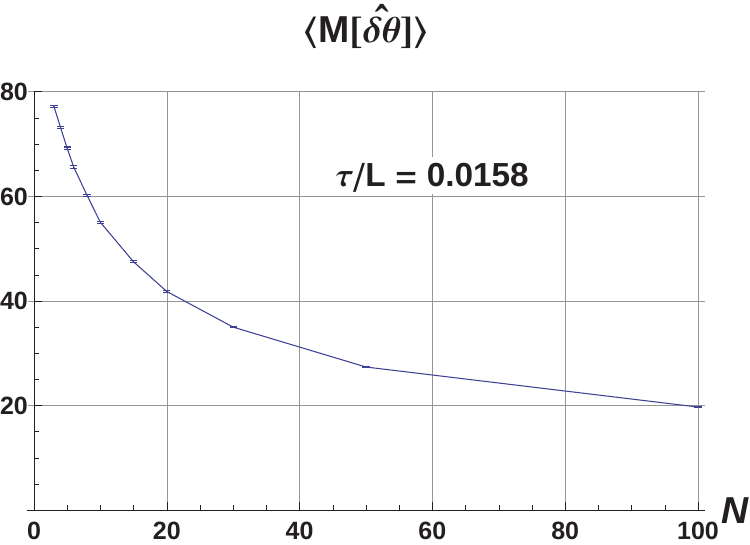}}
\spazio
  \centering
  \subfloat[Signed-distance $c$.]{\includegraphics[width=0.35\textwidth]{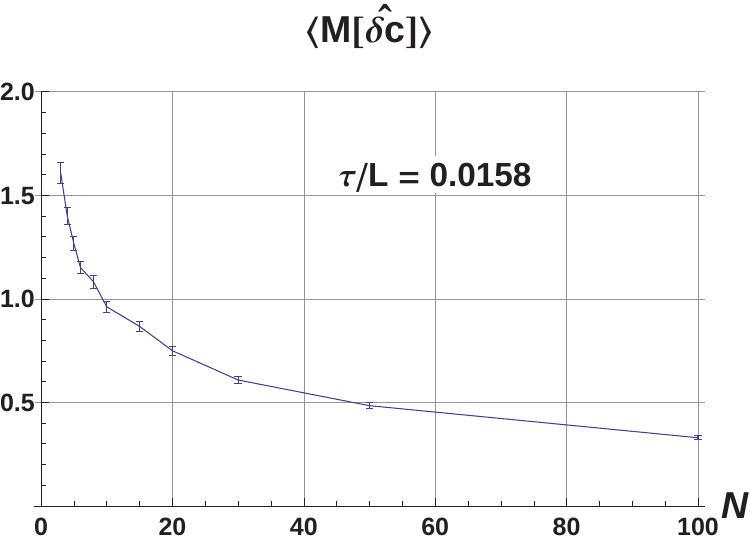}}
\end{center}  
\caption{Results of the simulations for $\av{M[\est{\ddd{\theta}}]}$ and $\av{M[\est{\ddd{c}}]}$,
         as a function of $N$,
         for $ \scaleFact/L = \pgra{ 0.0631; 0.0398; 0.0251; 0.0158 }$.}
\label{fig:ResMCMed2}
\end{figure}

\begin{figure}[htb] 
\begin{center}
  \centering
  \subfloat[   Angle $\theta$.  ]{\includegraphics[width=0.48\textwidth]{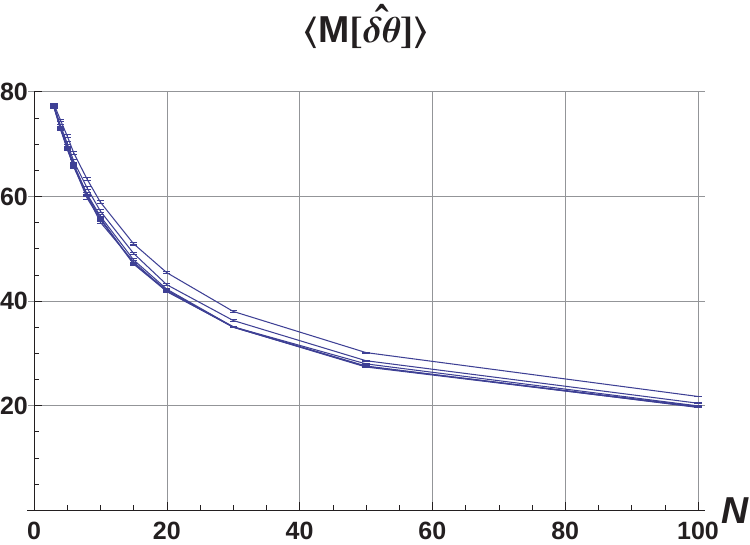}}
\spazio
  \centering
  \subfloat[Signed-distance $c$.]{\includegraphics[width=0.48\textwidth]{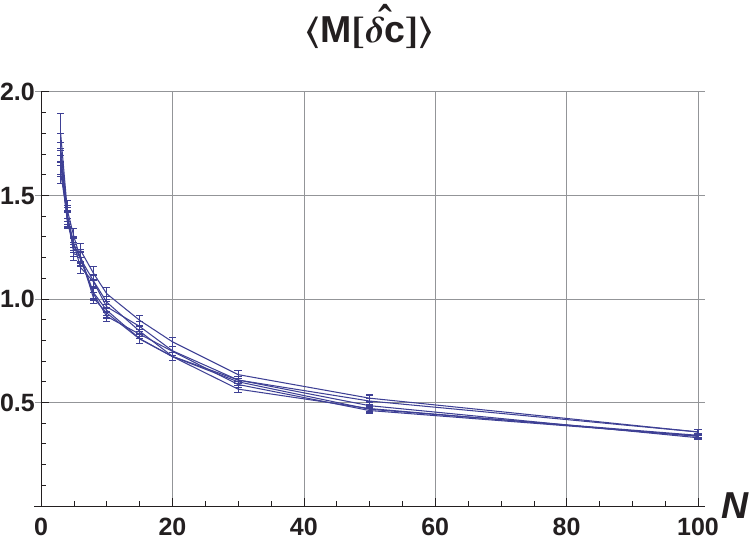}}
\end{center}  
\caption{Results of the simulations for $\av{M[\est{\ddd{\theta}}]}$ and $\av{M[\est{\ddd{c}}]}$,
         as a function of $N$,
         for $ \scaleFact/L = \pgra{ 0.1; 0.0631; 0.0398; 0.0251; 0.0158; 0.01 }$.
         Errors bars are very small and barely visible.}
\label{fig:ResMCMed2bis}
\end{figure}

Afterward,
in order to get rid of the variability associated with the random straight line
and random set of true data-points on the straight line, the values 
of $\av{M[\est{\ddd{\theta}}]}$ and $\av{M[\est{\ddd{c}}]}$
were normalized to $\av{\ddd{\theta}_0}$ and $\av{\ddd{c}_0}$.

For $\scaleFact \lesssim 0.02$, no statistically significant
difference with $\av{\ddd{\theta}_0}$ and $\av{\ddd{c}_0}$ has been found and the ratios 
$\av{M[\est{\ddd{\theta}}]}/\av{\ddd{\theta}_0}$ and $\av{M[\est{\ddd{c}}]}/\av{\ddd{c}_0}$ 
are compatible with one, as shown in Figure~\ref{fig:ResMCMed3}.

\begin{figure}[htb] 
\begin{center}
  \centering
  \subfloat[   Angle $\theta$.  ]{\includegraphics[width=0.48\textwidth]{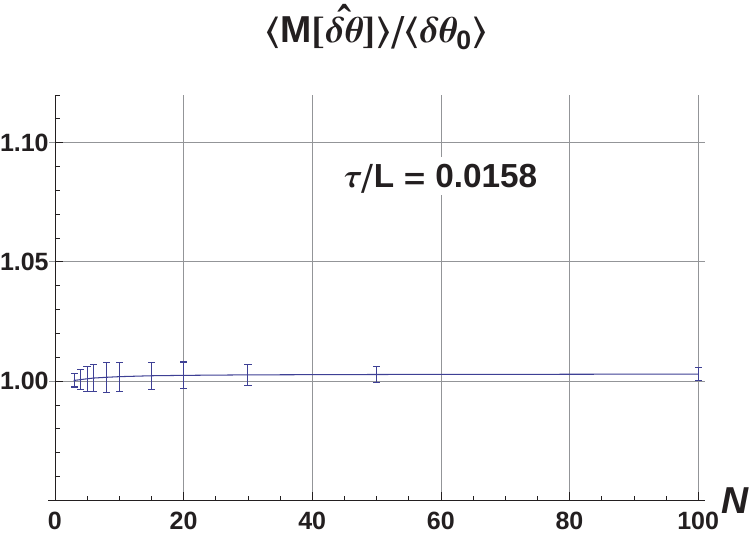}}
\spazio
  \centering
  \subfloat[Signed-distance $c$.]{\includegraphics[width=0.48\textwidth]{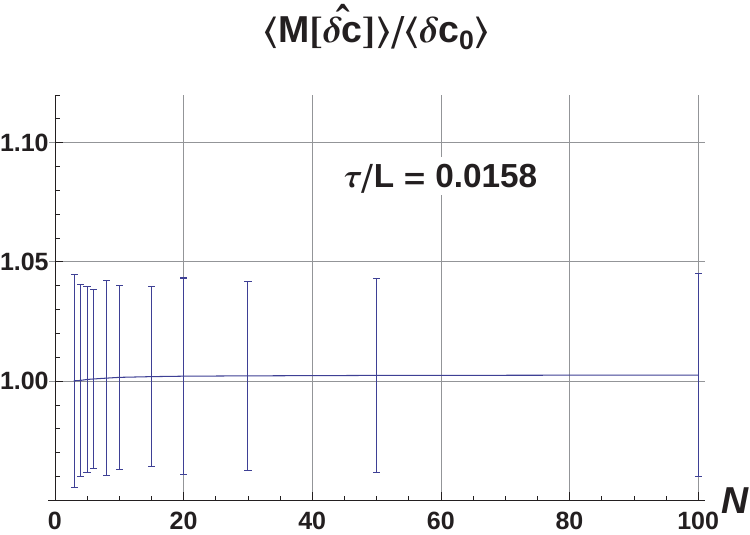}}
\end{center}  
\caption{Results of the simulations for the ratios $\av{M[\est{\ddd{\theta}}]}/\av{\ddd{\theta}_0}$ and $\av{M[\est{\ddd{c}}]}/\av{\ddd{c}_0}$,
         as a function of $N$,
         for $\scaleFact/L=0.0158$.}
\label{fig:ResMCMed3}
\end{figure}

On the other hand the behavior of the ratios 
$\av{M[\est{\ddd{\theta}}]}/\av{\ddd{\theta}_0}$ and $\av{M[\est{\ddd{c}}]}/\av{\ddd{c}_0}$ 
for values of $\scaleFact/L \gtrsim 0.02$,
as shown in Figure~\ref{fig:ResMCMed4}, 
shows a significant departure from the expected
$\av{\ddd{\theta}_0}$ and $\av{\ddd{c}_0}$, but not larger than $ \approx 10\%$ for $\scaleFact \lesssim 0.1$.

As a general trend, the ratios tend to one as $\scaleFact/L $ decreases, as
expected thanks to the improvement of the approximations made to derive the standard formula for the
propagation of errors.
Moreover the ratios tend to one at small $N$ anyway because at small $N$ the values of
the numerator and denominator became large with respect to their difference.

\begin{figure}[htb] 
\begin{center}
  \centering
  \subfloat[   Angle $\theta$.  ]{\includegraphics[width=0.48\textwidth]{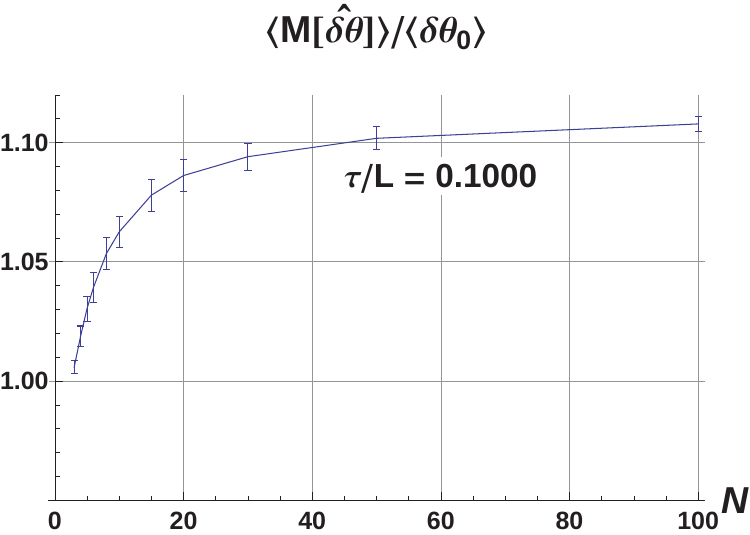}}
\spazio
  \centering
  \subfloat[Signed-distance $c$.]{\includegraphics[width=0.48\textwidth]{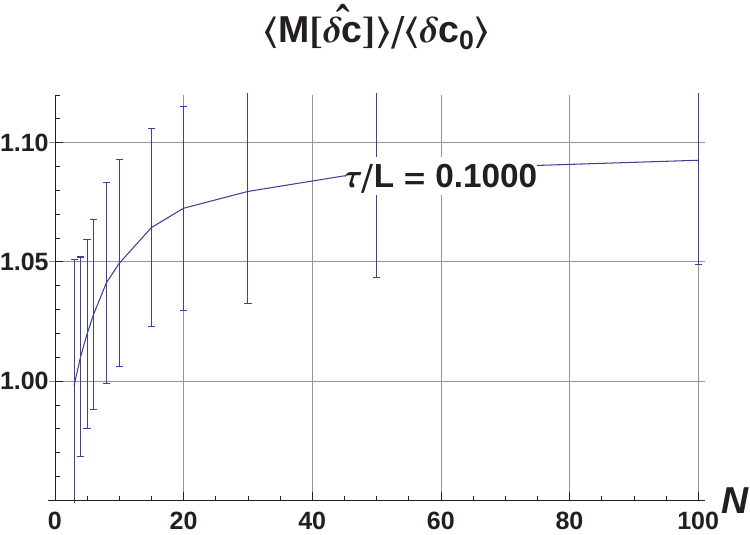}}
\end{center}  
\caption{Results of the simulations for the ratios $\av{M[\est{\ddd{\theta}}]}/\av{\ddd{\theta}_0}$ and $\av{M[\est{\ddd{c}}]}/\av{\ddd{c}_0}$,
         as a function of $N$,
         for $\scaleFact/L=0.1$.}
\label{fig:ResMCMed4}
\end{figure}

\clearpage

\subsection{Conclusions from the Monte Carlo simulations}

The main results of the Monte Carlo simulations presented in the previous
sections can be summarized as follows.

\begin{itemize}

\item
The standard errors, equations~\ref{eq:dtheta} and~\ref{eq:dc}, calculated from the true data points,
$\av{\ddd{(.)}_0}$,
scale as
$\sim 1/\sqrt{N+4}$, for $\theta$, and as
$\sim 1/\sqrt{N+1}$, for $c$, at fixed $\scaleFact/L$.
Normalization by $\scaleFact$ provides, for $\scaleFact/L \lesssim 0.02$,
a universal curve as a function of $N$.

\item
Both the standard deviation of the estimators of the parameters, $\theta$ and $c$, 
$\av{\est{\sigma}[\est{(.)}]}$,
and the standard errors, equations~\ref{eq:dtheta} and~\ref{eq:dc}, calculated from the measured data points,
$\av{M[\est{\ddd{(.)}}]}$,
as long as $\scaleFact/L \lesssim 0.05$, scale, at fixed $\scaleFact/L$, as
$\sim 1/\sqrt{N+4}$, for $\theta$, and as
$\sim 1/\sqrt{N+1}$, for $c$.
Normalization by $\scaleFact$ provides, for $\scaleFact/L \lesssim 0.02$,
a universal curve as a function of $N$.

\item
The standard errors, equations~\ref{eq:dtheta} and~\ref{eq:dc}, calculated from the measured data points,
$\av{M[\est{\ddd{\theta}}]}$ and $\av{M[\est{\ddd{c}}]}$,
are reliable estimates of the standard deviation of the estimators of $\theta$ and $c$,
$\av{\est{\sigma}[\est{(.)}]}$,
to within $\approx 10\%$ for $ \scaleFact / L \lesssim 0.1 $.
In particular, to within $\approx 10\%$, it shall not be necessary to re-evaluate the standard errors
using the estimated true data points values after the best-fit straight line has
been determined.

\item
For $ \scaleFact / L \lesssim 0.02 $, 
the standard errors, equations~\ref{eq:dtheta} and~\ref{eq:dc}, 
both calculated from the true data points, 
$\av{\ddd{(.)}_0}$,
and calculated from the measured data points, 
$\av{M[\est{\ddd{(.)}}]}$,
and 
the standard deviation of the estimators of the parameters,
$\av{\est{\sigma}[\est{(.)}]}$,
do not show any difference, within the statistical uncertainty of the Monte
Carlo simulations.

\end{itemize}

The following additional conclusions were obtained by the Monte Carlo
simulations, and stated without showing explicit evidence in this paper.

\begin{itemize}

\item
Lack of bias of the two estimators, to within the statistical uncertainty.

\item
Excellent normality of the distribution of the
$\est{\theta}$ estimator, for all the simulated parameters, to within the statistical
uncertainty, according to common statistical tests, such as Cram\'er-von Mises
and Kolmogorov-Smirnov~\cite{bi:James}.

\item
Excellent normality of the distribution of the
$\est{c}$ estimator, whenever $ \scaleFact/L \lesssim 0.01 $, to within the statistical
uncertainty. For larger values of $ \scaleFact/L $ strong deviations from normality start
to be very significant for small values of $N$ and less significant for large values
of $N$; for instance, Cram\'er-von Mises and Kolmogorov-Smirnov tests give a
P-value less than $ \approx 0.001$ 
whenever $ \scaleFact/L =0.1 $ and $N \lesssim 37$. 
The distributions, deviating from a Gaussian shape, show a positive excess kurtosis.

\end{itemize}


}

\pp{

\section{Toy Monte Carlo simulations}
\label{pa:simul}

In order to cross-check and evaluate the accuracy of the formulas for the standard errors
derived in this paper, extensive Monte Carlo
simulations~\footnote{\textit{ Mathematica, Version 8.0}, Wolfram Research, Inc., Champaign, IL (2010).}
were carried on, as described in the companion note~\cite{bi:nota}, which provides
full evidence for the following results.

\begin{itemize}

\item
The standard errors, equations~\ref{eq:dtheta} and~\ref{eq:dc}, calculated from the true data points,
scale as
$\sim 1/\sqrt{N+4}$, for $\theta$, and as
$\sim 1/\sqrt{N+1}$, for $c$, at fixed $\scaleFact/L$.
Normalization by $\scaleFact$ provides, for $\scaleFact/L \lesssim 0.02$,
a universal curve as a function of $N$.

\item
Both the standard deviation of the estimators of the parameters, $\theta$ and $c$, and
the standard errors, equations~\ref{eq:dtheta} and~\ref{eq:dc}, calculated from the measured data points,
as long as $\scaleFact/L \lesssim 0.05$, scale, at fixed $\scaleFact/L$, as
$\sim 1/\sqrt{N+4}$, for $\theta$, and as
$\sim 1/\sqrt{N+1}$, for $c$.
Normalization by $\scaleFact$ provides, for $\scaleFact/L \lesssim 0.02$,
a universal curve as a function of $N$.

\item
The standard errors, equations~\ref{eq:dtheta} and~\ref{eq:dc}, calculated from the measured data points,
are reliable estimates of the standard deviation of the estimators of $\theta$ and $c$,
to within $\approx 10\%$ for $ \scaleFact / L \lesssim 0.1 $.
In particular, to within $\approx 10\%$, it shall not be necessary to re-evaluate the standard errors
using the estimated true data points values after the best-fit straight line has
been determined.

\item
For $ \scaleFact / L \lesssim 0.02 $, 
the standard errors, equations~\ref{eq:dtheta} and~\ref{eq:dc}, 
both calculated from the true data points and calculated from the measured data points, and 
the standard deviation of the estimators of the parameters
do not show any difference, within the statistical uncertainty of the Monte
Carlo simulations.

\end{itemize}

}

\section{Conclusions}

Simple formulas for both the best-fit parameters and their standard errors in
the LSFSL-SWM have been derived, using the angle/signed-distance parametrization
of the straight line, and validated with Monte Carlo simulations.

Several properties of the standard errors were derived and investigated.
The standard errors in the slope/intercept parameterization were derived.
A simple relation for the case of highly-correlated measurements was derived.
The extent to which errors in one of the variables can be neglected is
quantified and the limiting case of the \mbox{\mbox{OLS-y:x}/\mbox{OLS-x:y}} fit
was studied.

\begin{acknowledgments}

The author thanks his two home Institutions, University of Genova and INFN (Italy) for support.

The constructive criticism and suggestions of two anonymous referees are gratefully acknowledged.

Finally, the author wishes to thank 
David Websdale (Imperial College, London, UK) and 
Olav Ullaland (CERN, European Laboratory for Particle Physics, Geneva, Switzerland)
for careful reading of the manuscript and for many useful suggestions.

\end{acknowledgments}

\clearpage

\appendix

\nn{

\section{Derivation of the best-fit line}
\label{app:ThetaAndsignedDistanceResults}

The search for stationary points of the error function in equation~\ref{eq:ChiSquare} (including the
minimum point) leads to the two equations:
\begin{gather}
        \label{eq:FirstDerivZeroTheta}
        \pton{ \DD{V} } \sin\pqua{ 2\theta } - 2 \CovXY \cos\pqua{ 2\theta } =0
        \\
        \label{eq:FirstDerivZeroC}
        \av{x} \sin\pqua{\theta} - \av{y} \cos\pqua{\theta} + c = 0
\punto
\end{gather}

The solution of equation~\ref{eq:FirstDerivZeroTheta} uses standard and
well-known trigonometric procedures. However some care is required for the
proper handling of the trigonometric functions.
After the angle $\theta$ is found, equation~\ref{eq:FirstDerivZeroC} allows to
determine $c$ in a trivial way.

Equation~\ref{eq:FirstDerivZeroTheta} always has two distinct solutions for
$2\theta$, differing by $\pm\pi$, in the interval $0 \leq 2\theta \lt 2\pi$, as
it is well-known from elementary trigonometry. In fact, letting 
$X= \cos\pqua{ 2\theta }$ and 
$Y= \sin\pqua{ 2\theta }$ and using $ X^2+Y^2=1$ 
the interpretation of the equation in terms of analytic geometry immediately leads to the conclusion.

Therefore there are four distinct solutions
for $\theta$, in the interval $0 \leq 2\theta \lt 2\pi$, differing by
$\pm\pi/2$. Two of them, differing by $\pm\pi$, minimize the error function and 
corresponds to the same straight line. The other two solutions, orthogonal to the first
ones, maximize the error function.

First note that the sign of the product $ \DD{V} \CovXY $ is the same as the sign
of the product $ \sin\pqua{ 2\theta } \cos\pqua{ 2\theta } $, so that this determines
in which quadrant the angle $ 2 \theta $ lies:
\begin{gather}
         \DD{V} \CovXY \lessgtr 0
         \ifoif
         \sin\pqua{ 2\theta } \cos\pqua{ 2\theta } \lessgtr 0
\punto
\end{gather}

Second, the following special cases arise:
\begin{gather}
         \DD{V} = 0
         \ifoif
         \cos\pqua{ 2\theta } = 0
         \ifoif
         \theta = \afrac{\pi}{4} + q\afrac{\pi}{2}
         \spazio q \in \mathbb{Z}
         \\
         \CovXY = 0
         \ifoif
         \sin\pqua{ 2\theta } = 0
         \ifoif
         \theta = q\afrac{\pi}{2}
         \spazio q \in \mathbb{Z}
\punto
\end{gather}

Finally, if $ \DD{V} \neq 0 $ equation~\ref{eq:FirstDerivZeroTheta} can be
safely reduced to:
\begin{gather}\label{eq:SolForTheta}
         \DD{V} \neq 0
         \ifoif
         \tan\pqua{ 2\theta } = \afrac{2 \CovXY }{ \DD{V} } 
         \spazio
         \theta = \afrac{1}{2}\arctan\pqua{\afrac{2 \CovXY }{ \DD{V} } } + q\afrac{\pi}{2}
         \spazio q \in \mathbb{Z}
\punto
\end{gather}

Equation~\ref{eq:SolForTheta} for $\theta$ provides two angles in the range 
$-\pi/2 \lt \theta \lt +\pi/2$, differing by $ \pi/2 $, one corresponding to a
minimum and the other to a maximum of the error function.

Equation~\ref{eq:SolForTheta} immediately implies the useful relations:
\begin{gather}
         \cos^2\pqua{ 2\theta } = 
         \afrac{ \pton{\DD{V}}^2 }{ \pton{\DD{V}}^2 + 4 \CovXY^2 }
         \spazio
         \sin^2\pqua{ 2\theta } = 
         \afrac{  4 \CovXY^2 }{ \pton{\DD{V}}^2 + 4 \CovXY^2 }
\punto
\end{gather}


In order to determine which solution in equation~\ref{eq:SolForTheta} corresponds to the minimum/maximum it is
easiest to re-start from equation~\ref{eq:ChiSquare}, and re-write it as
follows, using a few trigonometric transformations:
\begin{gather}
    \label{eq:DistToQuadForm1}
    \afrac{ \scaleFact^2 \chiSq[\theta,c] }{N} \equiv
    \afrac{1}{N} \sum_{k=1}^N \pton{ x_k \sin\pqua{\theta} - y_k \cos\pqua{\theta} + c }^2 =
    \\
    \label{eq:DistToQuadForm2}
    = \afrac{1}{N} \sum_{k=1}^N \pton{ \pton{ x_k - \av{x} } \sin\pqua{\theta} - \pton{ y_k - \av{y} } \cos\pqua{\theta} }^2 =
    \\
    \label{eq:DistToQuadForm3}
    = \VarX \sin^2\pqua{\theta} + \VarY \cos^2\pqua{\theta} - 2 \CovXY \sin\pqua{\theta}\cos\pqua{\theta} =
    \\
    = \afrac{1}{2}\pton{ \pton{ \VarX + \VarY } - \DD{V} \cos\pqua{2\theta} - 2 \CovXY \sin\pqua{2\theta} } =
    \\
    = \afrac{1}{2}\pton{ \pton{ \VarX + \VarY } - \cos\pqua{2\theta} \afrac{ \pton{\DD{V}}^2 + 4 \CovXY^2 }{\DD{V}} } =
    \spazio\text{(for $\DD{V} \neq 0$)}
    \\
    = \afrac{1}{2}\pton{ \pton{ \VarX + \VarY } - \sin\pqua{2\theta} \afrac{ \pton{\DD{V}}^2 + 4 \CovXY^2 }{2\CovXY} } =
    \spazio\text{(for $\CovXY \neq 0$)}
    \\
    \label{eq:DistToQuadForm4}
    \arrforw \afrac{1}{2}\pton{ \pton{ \VarX + \VarY } - \sqrt{ \pton{\DD{V}}^2 + 4 \CovXY^2 } }
    \spazio\text{at the minimum} 
    \punto
\end{gather}

The above expressions clearly show that the minimum is obtained for those values
of $\theta$ such that 
$ \DD{V} \cos\pqua{2\theta} > 0 $ and 
$ \CovXY \sin\pqua{2\theta} > 0 $. Therefore:
\begin{gather}
        \DD{V} \cos\pqua{2\theta} > 0
        \arrforw
        \cos\pqua{2\theta} = \afrac{\DD{V}}{ \sqrt{ \pton{\DD{V}}^2 + 4 \CovXY^2
        } }
\virgola
        \\
        \CovXY \sin\pqua{2\theta} > 0
        \arrforw
        \sin\pqua{2\theta} = \afrac{ 2 \CovXY }{ \sqrt{ \pton{\DD{V}}^2 + 4 \CovXY^2 } }
\punto
\end{gather}

The above equations show that the angle $2\theta$ corresponding to the best-fit
line lies in the first/fourth quadrant according to the positive/negative sign
of $\CovXY$.

Clearly, the minimum value of the error function is zero for a perfect line fit:
$ \CovXY^2 = \VarX \VarY $.

The minimum of the error function can finally be expressed, in terms of 
$ \alpha \equiv \sign\pqua{\CovXY} \equiv \CovXY / \abs{\CovXY}$ which
makes sure that $\sin\pqua{2\theta}$ has the correct sign, as:
\begin{gather}
        \theta = \afrac{\alpha}{2} \arccos{ \pton{\afrac{\DD{V}}{ \sqrt{ \pton{\DD{V}}^2 + 4 \CovXY^2} } } }
        + q \pi \spazio q \in \mathbb{Z}
        \spazio\text{choosing $\abs{\theta} \lt \pi/2$}
\punto
\end{gather}

Alternatively the minimum/maximum can be determined as follows.
Taking the second derivative with respect to $\theta$ of the
equation~\ref{eq:DistToQuadForm3}, after using the stationary point conditions for
both $c$ and $\theta$, one finds:
\begin{gather}
    \derivsq{}{\theta}
    \pton{ \afrac{ \scaleFact^2 \chiSq[\theta] }{N} } =
    2 \DD{V} \cos\pqua{2\theta} + 4 \CovXY \sin\pqua{2\theta} =
    2\afrac{ \sin\pqua{2\theta} }{2\CovXY} \pton{ \pton{\DD{V}}^2 + 4\CovXY^2 }=
    2\afrac{ \cos\pqua{2\theta} }{\DD{V} } \pton{ \pton{\DD{V}}^2 + 4\CovXY^2 }
\virgola
\end{gather}
showing again that the minimum is given by those values of $\theta$ such that 
$\sin\pqua{2\theta}$ and $\CovXY$ have the same sign.

}

\nn{

\section{Limit of the \texorpdfstring{\mbox{\mbox{OLS-y:x}/\mbox{OLS-x:y}}}{.....} fit}
\label{app:LimOLS}

The limiting cases of the \mbox{\mbox{OLS-y:x}/\mbox{OLS-x:y}} fit, that is the case
when one of the two variables has negligible errors, can be recovered.
Consider a fixed set of $N$ measured data points, 
$ \pgra{ \tz{x}_k \pm \StDvXt ; \tz{y}_k \pm \StDvYt }, ( k = 1, \ldots, N )$, and let us study the
limiting case when one of the standard errors tends to zero.

The relation between the raw and re-scaled slopes is:
\begin{gather}
        \tan\pqua{\tz{\theta}} = \afrac{\StDvYt}{\StDvXt} \tan\pqua{\theta}
\punto
\end{gather}

\subsection{Limit for the slope/intercept}

Consider, for definiteness, the case \mbox{OLS-y:x}.

Consider first the relations for the slope.

In the limit $\StDvXt \longrightarrow 0 $, 
for equation~\ref{eq:SlopeInterceptYSol}, one finds,
the expression for the slope of the \mbox{OLS-y:x} fit as follows:
\begin{gather}
        p_{\tz{y}} \equiv \tan\pqua{\thtx} =
        \afrac{\StDvYt}{\StDvXt} \pton{ A_y + \alpha \sqrt{1+A_y^2} } 
        \spazio
        A_y \equiv \afrac{ \StDvXt^2 \VarYt - \StDvYt^2 \VarXt }{ 2 \StDvXt\StDvYt\CovXYt }
        \\
        \StDvXt \longrightarrow 0 \arrforw  
        A_y \approx \afrac{ - \StDvYt \VarXt }{ 2 \StDvXt\CovXYt }
        \arrforw
        p_{\tz{y}} \approx 
        \afrac{\StDvYt}{\StDvXt} \pton{ A_y + \alpha |A_y | \pton{ 1 + \afrac{1}{2 A_y^2} } } \longrightarrow
        \afrac{ \CovXYt  } { \VarXt }
\punto
\end{gather}

Consider now the relations for the intercepts:
the equation in~\ref{eq:SlopeInterceptYSol} for $q$ is the same as the
one for \mbox{OLS-y:x} fit, so there is need to investigate further its limit.

Similarly, one may proceed for the limiting case $\StDvYt \longrightarrow 0 $, to
find the \mbox{OLS-x:y} fit limit.

\subsection{Limit for the standard errors}

First re-write equation~\ref{eq:dtheta} in
terms of the raw variables:
\begin{gather}
        \ddd{\theta} = \afrac{1}{\sqrt{N}}\sqrt{ 
        \afrac{ \StDvXt^2 \StDvYt^2 \pton{ \StDvYt^2 \VarXt + \StDvXt^2 \VarYt } }
              { \pton{\StDvYt^2\VarXt - \StDvXt^2\VarYt}^2 + 4 \StDvXt^2 \StDvYt^2\CovXYt^2 }
}
\punto
\end{gather}

The identification of the angle $\theta$ with the angles in
equation~\ref{eq:SlopeIntercept} (see also equation~\ref{eq:AngsDefs}) is obviously:
\begin{gather}
        \tz{\theta}_y \equiv \tz{\theta}
        \spazio
        \tz{\theta}_x \equiv \afrac{\pi}{2} - \tz{\theta}
        \arrforw
        \abs{ \ddd{ \tz{\theta} } } =
        \abs{ \ddd{ \tz{\theta}_y } } =
        \abs{ \ddd{ \tz{\theta}_x } }
\punto
\end{gather}


The relation between the error on the raw angle and the re-scaled angle is then:
\begin{gather}\label{eq:TheUsefulEquationForOLSLimit}
        \ddd{ \tz{\theta} } 
        = 
        \afrac{\StDvYt}{\StDvXt}\pton{\afrac{ 1 + \tan^2\pqua{\theta} }{ 1 + \tan^2\pqua{\tz{\theta}} }}  
        \ddd{ \theta } 
        =
        \afrac{ \StDvYt^2 + \StDvXt^2 \tan^2\pqua{\tz{\theta}} }{ 1 + \tan^2\pqua{\tz{\theta}} } 
        \afrac{ 1 }{\StDvXt\StDvYt} 
        \ddd{ \theta }
        =
        \afrac{ \StDvYt^2 \cos^2\pqua{\tz{\theta}} + \StDvXt^2 \sin^2\pqua{\tz{\theta}} }{\StDvXt\StDvYt} 
        \ddd{ \theta }
\punto
\end{gather}

From the above relation one can find for the raw angle:
\begin{gather}
        \begin{cases}
                \StDvX \longrightarrow 0 \arrforw 
                \ddd{ \tz{\theta} } \sim \afrac{\StDvYt}{\StDvXt} \cos^2\pqua{\tz{\theta}} \ddd{ \theta }
                \\
                \StDvY \longrightarrow 0 \arrforw 
                \ddd{ \tz{\theta} } \sim \afrac{\StDvXt}{\StDvYt} \sin^2\pqua{\tz{\theta}} \ddd{ \theta }
        \end{cases}
\punto
\end{gather}
 
The above relation gives the error on the raw slope:
\begin{gather}
        \begin{cases}
                \StDvX \longrightarrow 0 \arrforw 
                \ddd{ p_{\tz{y}} } =
                \pton{ 1 + \tan^2\pqua{\tz{\theta}_y} } \ddd{ \tz{\theta}_y } \sim
                \afrac{\StDvYt}{\StDvXt} \ddd{ \theta }
                \sim
                \afrac{1}{\sqrt{N}} \afrac{ \StDvYt }{\sqrt{ \VarXt }}
                \\
                \StDvY \longrightarrow 0 
                \arrforw 
                \ddd{p_{\tz{x}}} = 
                \pton{ 1 + \tan^2\pqua{\tz{\theta}_x} } \ddd{ \tz{\theta}_x }=
                \afrac{\StDvXt}{\StDvYt} \ddd{ \theta }\sim
                \afrac{1}{\sqrt{N}} \afrac{ \StDvXt }{\sqrt{ \VarYt }}
%
%
        \end{cases}
\virgola
\end{gather}
exactly the ones for the \mbox{\mbox{OLS-y:x}/\mbox{OLS-x:y}} fit.

Again, as the equations for the intercepts are the same as for the 
\mbox{\mbox{OLS-y:x}/\mbox{OLS-x:y}} fit, there is no need to investigate the
limiting case.

}

\nn{

\section{Derivation of the standard errors for slope and intercept}
\label{app:ErrorsSlopeIntercept}

The errors on the slope and intercept (see section~\ref{pa:SlopeIntercept}) can be easily calculated as a by-product
of the results obtained for the angle/signed-distance parametrization, starting
from:
\begin{gather}
        p[\theta,c] \equiv p_y = 
        \tan\pqua{\theta}
        \spazio
        q[\theta,c] \equiv q_y = 
        c \sqrt{ 1 + \tan^2\pqua{\theta} } =
        c / \cos\pqua{\theta} 
        \spazio
        \cos\pqua{\theta} \geq 0
\virgola
\end{gather}
without neglecting the covariance term in equation~\ref{eq:dctheta}.
Note that the sign of $c$ is the same as the sign of $q_y$.

\begin{gather}
        \label{eq:ErrorSlope}
        \pton{\ddd{p}}^2 = \pton{ 1 + p^2 }^2 \pton{\ddd{\theta}}^2
        \\
        \label{eq:ErrorIntercept}
        \pton{\ddd{q}}^2 = 
        \pton{ 1 + p^2 } \pton{ \afrac{\scaleFact^2}{N} + \pton{\ddd{\theta}}^2 \pton{Z - c p }^2 } =
        \pton{ 1 + p^2 } \pton{ \afrac{\scaleFact^2}{N} + \pton{\ddd{\theta}}^2 \av{x}^2 \pton{ 1 + p^2 } }
        \\
        \label{eq:CovPQ}
        \covariance\pqua{p,q} = 
        - \pton{ 1 + p^2 }^{3/2} \pton{Z - c p } \pton{\ddd{\theta}}^2 =
        - \pton{ 1 + p^2 }^2 \av{x} \pton{\ddd{\theta}}^2
\punto
\end{gather}

Equation~\ref{eq:ErrorIntercept} has a simple interpretation by observing that 
$ \pton{Z - c p }^2 = \av{x}^2 \pton{ 1 + p^2 } $ is the squared-distance between the centroid of the measured data points
and the point where the best-fit line crosses the $y$ axis.
As the best-fit line always passes by the centroid, the term 
$ \pton{ 1 + p^2 } \pton{\ddd{\theta}}^2 \pton{Z - c p }^2 $ is the
uncertainty on the intercept caused by the uncertainty of the angle, 
$ \pton{\ddd{\theta}}^2 $.
It is summed in quadrature to the term giving the error on the location of the
centroid.

It can be easily shown, by using the results of appendix~\ref{app:LimOLS}, in
particular equation~\ref{eq:TheUsefulEquationForOLSLimit}, that the above
formulas~\ref{eq:ErrorSlope},~\ref{eq:ErrorIntercept} and~\ref{eq:CovPQ}, tend
to the well-known results for the OLS fit.

}


\end{document}